\documentclass{aa} 
\bibpunct{(}{)}{;}{a}{}{,}

\usepackage{graphicx}
\usepackage{txfonts}
\usepackage{mathrsfs}
\usepackage{textcomp}
\usepackage{amssymb,amsmath,bm,graphicx}
\usepackage{stfloats}
\usepackage{caption}
\usepackage{fixltx2e}

\newcommand{\abs}[1]{\left|#1\right|}

\newcommand{\Mc}[1]{{\mathcal #1}}
\newcommand{\Eq}[1]{Eq.~(\ref{eq:#1})}
\newcommand{\Eqs}[2]{Eqs.~(\ref{eq:#1},~\ref{eq:#2})}
\newcommand{\Eqss}[3]{Eqs.~(\ref{eq:#1}),~(\ref{eq:#2}) and~(\ref{eq:#3})}
\newcommand{\Fig}[1]{Figure~\ref{fig:#1}}

\newcommand{\Tab}[1]{Table~\ref{tab:#1}}

\newcommand{\Sec}[1]{Section~\ref{sect:#1}}
\newcommand{\Secs}[2]{Sections~\ref{sect:#1} and~\ref{sect:#2}}
\newcommand{\App}[1]{Appendix~\ref{sect:#1}}

\newcommand{\Cov}{\textrm{Cov}}
\newcommand{\Var}{\textrm{Var}}
\newcommand{\xv}[0]{\mathbf{x}}
\newcommand{\kv}[0]{\mathbf{k}}

\newcommand{\bx}[0]{\mathbf{x}}
\newcommand{\Mcc}[0]{\mathscr{C}}
\newcommand{\Ecc}[0]{\overline{C}}
\newcommand{\Et}[0]{\overline{\tau}}
\newcommand{\refereeFirst}[1]{{#1}}
\newcommand{\refereeSecond}[1]{{#1}}
\DeclareMathOperator{\sgn}{sgn}

\DeclareMathOperator{\Win}{Win}

\newcommand{\Nset}{\mathbb{N}}
\newcommand{\Zset}{\mathbb{Z}}

\newcommand{\calN}{\mathcal{N}}

\newcommand{\paren}[1]{\left( #1\right)}
\newcommand{\bracket}[1]{\left[ #1\right]}

\newcommand{\EW}[1]{\mathbb{E}\left[#1\right]}
\newcommand{\HT}[0]{h_t}
\newcommand{\NT}[0]{N}

\begin{document}

   \title{Generalization of the noise model for time-distance helioseismology}


   \author{D. Fournier
          \inst{1}
          \and
          L. Gizon \inst{2,3}
          \and
          T. Hohage \inst{1}
          \and
          {A.~C.~Birch \inst{2}}
          }

   \institute{Institut f\"ur Numerische und Angewandte Mathematik,
              Lotzestrasse 16-18, 37083 G\"ottingen, Germany \\
              \email{d.fournier@math.uni-goettingen.de}
         \and
            Max-Planck-Institut f\"ur Sonnensystemforschung, Justus-von-Liebig-Weg 3, 37077 G\"ottingen, Germany
             \and
             Institut f\"ur Astrophysik, Georg-August-Universit\"at G\"ottingen,  Friedrich-Hund-Platz 1, 37077 G\"ottingen, Germany
             }

   \date{Received ; accepted }

 
  \abstract
   {In time-distance helioseismology, information about the solar interior is encoded in measurements of travel times between pairs of points on the solar surface. Travel times are deduced from the cross-covariance of the random wave field. Here we consider travel times and also products of travel times as observables. They contain information about e.g. the statistical properties of convection in the Sun.}
   {Using the travel time definition of \citet{GIZ04} we derive analytic formulae for the noise covariance matrix of travel times and products of travel times.}
   {The basic assumption of the model is that noise is the result of the stochastic excitation of solar waves, a random process which is stationary and Gaussian. We generalize the existing noise model by dropping the assumption of horizontal spatial homogeneity. Using a recurrence relation, we calculate the noise covariance matrices for the moments of order 4, 6, and 8 of the observed wave field, for the moments of order 2, 3 and 4 of the cross-covariance, and for the moments of order 2, 3 and 4 of the travel times.}
   { All noise covariance matrices depend only on the expectation value of the cross-covariance of the observed wave field.  For products of travel times, the noise covariance matrix consists of three terms proportional to $1/T$, $1/T^2$, and $1/T^3$, where $T$ is the duration of the observations. For typical observation times of a few hours, the term proportional to $1/T^2$ dominates and $\Cov[\tau_1 \tau_2, \tau_3 \tau_4] \approx \Cov[\tau_1, \tau_3] \Cov[\tau_2, \tau_4] + \Cov[\tau_1, \tau_4] \Cov[\tau_2, \tau_3]$, where the $\tau_i$ are arbitrary travel times. This result is confirmed for $p_1$ travel times by Monte Carlo simulations and comparisons with SDO/HMI observations.}
   {General and accurate formulae have been derived to model the noise covariance matrix of helioseismic travel times and products of travel times. These results could easily be generalized to other methods of local helioseismology, such as helioseismic holography and ring diagram analysis.}


   \maketitle

%

\section{Introduction}
The purpose of time-distance helioseismology \citep[][and references therein]{DUV93,GIZ05}  is to infer the subsurface structure and dynamics of the Sun using spatial-temporal correlations of the random wave field observed at the solar surface. Wave travel times between pairs of points (denoted $\tau$) are measured from the cross-covariance function. Wave speed perturbations and vector flows are then obtained by inversion of the travel times (e.g. \citet{KOS96, JAC12}). Such inversions require knowledge of the noise covariance matrix Cov[$\tau, \tau$]. Typically, noise is very high and strong correlations exist among travel times. \citet{GIZ04} studied the noise properties of travel times and derived a simple noise model that successfully explains the observations. The model is based on the assumption that the stochastic noise is stationary and horizontally spatially homogeneous, as a result of the excitation of waves by turbulent convection. In addition to time-distance helioseismology, this noise model has 
found applications in direct modeling inversions \citep{WOO06, WOO09} and ring-diagram analysis \citep{BIR07b}.

Time-distance helioseismology has been successfully applied to map flow velocities, $v_j$,  at supergranulation scales \citep[]{KOS96, DUV00, GIZ01, JAC08}. The statistical properties of convection can further be studied by computing horizontal averages of the turbulent velocities. For example, \citet{DUV00, GIZ10} showed that the horizontal divergence and the vertical vorticity of the flows are correlated through the influence of the Coriolis force on convection. It would be highly desirable to extract additional properties of the turbulent velocities, for example the (anisotropic) Reynolds stresses $\langle v_i v_j \rangle$ that control the global dynamics of the Sun \citep[differential rotation and meridional circulation, see][]{KIT05}. The noise associated with such measurement involves the fourth order moments of the travel times, $\Cov[\tau \tau, \tau \tau]$.

Alternatively, we would like to consider spatial averages of products of travel times $\langle \tau \tau \rangle$ as the fundamental data from which to infer the Reynolds stresses (or other second-order moments of turbulence). Spatial averages are meaningful when turbulent flows are horizontally homogeneous over the averaging region.  Inversions of average products of travel times are desirable since input data are fewer and less noisy. Once again, we need to know the noise covariance matrix $\Cov[\langle \tau \tau \rangle, \langle \tau \tau \rangle]$ in order to perform the inversion. 

In this paper, we study the noise properties of travel times and products of travel times. In \Sec{data}, the definitions for the cross-covariance function and the travel times are given. \Sec{noiseFormula} presents the assumptions of the noise model generalizing the model of \citet{GIZ04}. In \Sec{noiseTau} and in the Appendix, we derive analytical formulae for the noise covariance matrices of travel times and products of travel times. These formulae are confirmed in \Sec{noiseNum} by comparison to numerical Monte Carlo simulations and to SDO/HMI observations. The effects of horizontal spatial averaging are considered in \Sec{averages}.


%

\section{Observables: cross-covariance function, travel times, and products of travel times} \label{sect:data}

The fundamental observation in helioseismology is the filtered line-of-sight Doppler velocity $\phi(\xv, t)$ at points $\xv$ on the surface of the Sun 
and at times $t$. The filter acts by multiplication in the Fourier domain. In this paper we will only consider the $p_1-$ridge filter as an example.  Note that all the results presented in this paper do not depend on the choice of the filter. The signal $\phi(\xv, t)$ is recorded over a duration time $T=(2\NT+1)\HT$ where 
$\HT$ is the temporal resolution at observation times $t_n = n\HT$ for $n=-\NT,\dots,\NT$. The observed wavefield during the observation time $T$ is denoted $\phi_T$. We have $\phi_T(\xv, t) = \phi(\xv, t) \Win_T(t)$  where $\Win_T$ is a window function (equal to 1 if $\abs{t} \leq T/2$ and 0 otherwise).

Helioseismic analysis is performed in Fourier space. Let us define the temporal Fourier transform of $\phi_T$ by
\[\phi_T(\xv,\omega):= \frac{\HT}{2\pi}\sum_{n=-\NT}^{\NT} \phi(\xv, t_n) \exp(i\omega t_n).
\]
The frequencies $\omega$ are treated as continuous variables in the remainder of this paper in order to be able to take into account the frequency correlations (see \Sec{freqCorr}).
The cross-covariance function between two points at the surface of the Sun is a multiplication in the Fourier domain \citep{DUV93}
\begin{equation}
C(\xv_1, \xv_2, \omega) 
= \frac{2\pi}{T}\phi_T^\ast(\xv_1, \omega) \phi_T(\xv_2, \omega). \label{eq:Cfourier}
\end{equation}
Working in Fourier space is faster (and easier). 
In the time-domain the cross-covariance becomes 
\begin{equation}\label{eq:simple_cov_estim}
C(\xv_1, \xv_2, t_n) 
:=\frac{1}{2\NT+1}\sum_{j=\max(-\NT,-\NT-n)}^{\min(\NT,\NT-n)}
\phi(\xv_1,t_j)\phi(\xv_2,t_{j+n}) .
\end{equation}
where $t_n$ is the correlation time lag.

%
%
%
%

Cross-covariances are the basic data to compute the travel times. We denote $\tau_{+}(\xv_1, \xv_2)$  the travel time for a wave packet traveling from point $\xv_1$ to point $\xv_2$ and $\tau_{-}(\xv_1, \xv_2)$ the travel time for a wave packet traveling from $\xv_2$ to $\xv_1$. In the limit discussed by \citet{GIZ04} the \refereeFirst{incremental} travel times can be measured from the estimated cross-covariance using
\begin{align}
 \tau_\pm(\xv_1, \xv_2) := \HT \sum_{n=-\NT}^{\NT} &W_\pm(\xv_1, \xv_2, t_n) \times \nonumber \\
 &\left( C(\xv_1, \xv_2, t_n) - C^{\textrm{ref}}(\xv_1, \xv_2, t_n) \right) \label{eq:taudef}
\end{align}
where $C^{\textrm{ref}}$ is a deterministic reference cross-covariance coming from spatial averaging or from a solar model and the weight function $W_\pm$ are defined as
\begin{equation}
 W_\pm(\xv_1, \xv_2, t) := \frac{\mp f(\pm t) \partial_t C^{\textrm{ref}}(\xv_1, \xv_2, t)}{\HT \sum_n f(\pm t_n) [\partial_t C^{\textrm{ref}}(\xv_1, \xv_2, t_n)]^2} \label{eq:W}
\end{equation}
with $f$ a window function used to select an interval of time around the first arrival time of the wave packet (for example, a cut-off function). 
Notice that for spatially homogeneous noise we generally choose that $C^{\textrm{ref}}(\xv_1, \xv_2, t) = C^{\textrm{ref}}(\xv_2 - \xv_1, t)$ which implies that $W(\xv_1, \xv_2, t) = W(\xv_2 - \xv_1, t)$. However, this assumption is not necessary in the remainder of this paper.

We write $\tau_\alpha$ where the subscript $$\alpha \in \{ +, -, \textrm{diff}, \textrm{mean} \}$$ denotes the type of travel time and the corresponding weight function $W_\alpha$. The mean and difference travel times $\tau_{\textrm{diff}}$ and $\tau_{\textrm{mean}}$ can be obtained from the one way travel times by $\tau_{\textrm{diff}} = \tau_+ - \tau_-$ and $\tau_{\textrm{mean}} = (\tau_+ + \tau_-)/2$.

In this paper, we are interested in the noise covariance matrix for travel times $\tau_{\alpha_1}(\xv_1, \xv_2)$ and
products of travel times $\tau_{\alpha_1}(\xv_1, \xv_2) 
\tau_{\alpha_2}(\xv_3, \xv_4)$  where $\tau$ is defined by 
\Eq{taudef}. To simplify the notations, let
\begin{equation*}
 \tau_1 := \tau_{\alpha_1}(\xv_1, \xv_2), \quad  \tau_2 := \tau_{\alpha_2}(\xv_3, \xv_4),
\end{equation*}
\begin{equation}
 \textrm{and more generally } \tau_i := \tau_{\alpha_i}(\xv_{2i-1}, \xv_{2i}). \label{eq:taudefinition}
\end{equation}

\section{Generalization of the noise model} \label{sect:noiseFormula}

\subsection{Assumptions} \label{sect:asumptions}

\begin{figure}
  \includegraphics[width=0.9\linewidth]{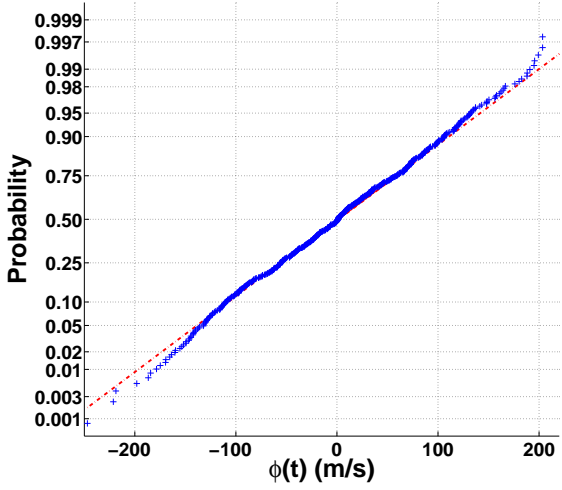}
 \caption{Probability density plot representing the filtered line-of-sight velocity $\phi(t)$ for a $p_1-$ridge with an observation time $T=8$ h. For Gaussian observations, all data should be on a straight line.} \label{fig:normalProbabilityPlot}
\end{figure}

The basic assumption of the noise model is the following: The observations at the relevant spatial points $\xv_1,\dots,\xv_M$ are described by a vector-valued stationary Gaussian time series $\mathbf{(\phi(\xv_1,t_n),\dots,\phi(\xv_M,t_n))}$. For the sake of simplicity, we can also assume without loss of generality that $\EW{\phi(\xv_m, t_n)}=0$ at each $\xv_m$ for all $n\in\Zset$.  This model is valid in the quiet Sun (away from evolving active regions) but does not assume that the noise is spatially homogeneous contrary to the model of \citet{GIZ04} as detailed in \Sec{comparisonGB}.  \refereeSecond{This assumption is supported by the observed distribution of the HMI Doppler velocity: \Fig{normalProbabilityPlot} shows the probability density of the filtered line-of-sight velocity. For a Gaussian distribution, the data should line up along a straight line. We can see a very good agreement for probabilities betweeen $5\%$ and $95\%$. The deviations in the tail of the plot (for probabilities smaller than $5\%$) may be due to statistical errors (as we have less realisations for these events).}

One may also replace the spatial points by some spatial averages. Such averages are often used to improve the signal-to-noise ratio
We will denote $\Ecc$ the expectation value of the cross-covariance
\begin{equation}
 \Ecc(\xv_a,\xv_b,\omega) = \EW{C(\xv_a,\xv_b,\omega)} = \frac{2\pi}{T}\EW{\phi_T^\ast(\xv_a,\omega)\phi_T(\xv_b,\omega)}.
\end{equation}


\subsection{Independance of the geometry} \label{sect:comparisonGB}

\citet{GIZ04} assumed that the observations $\phi(\xv_{ij},t_n)$ are given on a Cartesian grid $\{\xv_{ij}\}$ by an approximately flat patch of the Sun's surface. The discrete Fourier transform of the finite dimensional signal was assumed to be of the form
\begin{equation}\label{eq:noise_model_GB}
\phi({\bf k}_{ij},\omega_l) 
= \sqrt{\mathcal{P}({\bf k}_{ij},\omega_l)}\calN_{ijl}
\end{equation}
where $\Mc{P}$ is the power spectrum, $\omega_l:=2\pi l/T$, 
 and $\calN_{ijl}$ are complex independent and identically distributed Gaussian variables with zero-mean and unit variance. In this case, the frequency correlations were ignored and
\begin{equation}\label{eq:GB_model}
\frac{2\pi}{T}\EW{\phi^{\ast}(\xv_a,\omega_j) \phi(\xv_b,\omega_l)}
= \delta_{jl}\Ecc_{GB}(\xv_b-\xv_a,\omega_l)
\end{equation}
was assumed. We have denoted $\Ecc_{GB}$ the expectation value of the cross-covariance used by \citet{GIZ04}. Our assumption is more general as it does not require a planar geometry and allows a natural treatment of spatially averaged quantities. It means that all our results are valid in any geometry and it is in particular the case for the results presented in \citet{GIZ04}.

\subsection{On frequency correlations} \label{sect:freqCorr}
As the observation time $T$ is finite, the discrete Fourier transforms $\phi_T(\xv, \omega_j)$ and $\phi_T(\xv, \omega_l)$ for $j\neq l$ are no longer uncorrelated because of the window function. The necessity of a correction term for finite $T$ was discussed, but not further analyzed in \citet{GIZ04}. It turns out that there is an explicit formula for this correction term in terms of the periodic Hilbert transform of $\Ecc$ and a smoothed version of $\Ecc$. The exact formulation is given in \App{convergenceApp} where it is also shown that the error made by considering a finite observation time can be bounded 
\begin{align}
&\sup_{j, l}\left|\frac{2\pi}{T}\EW{\phi_T^{\ast}(\xv_a,\omega_j)\phi_T(\xv_b,\omega_l)} - \delta_{jl}\Ecc_{GB}(\xv_b-\xv_a,\omega_l)]\right| \nonumber \\
&\leq \frac{\HT}{4T} \abs{\sum_{k=-2\NT}^{2\NT} \abs{t_k} \Ecc(\xv_a, \xv_b, t_k)}. \label{eq:error_bound}
\end{align}
Note that the right hand side of \eqref{eq:error_bound} depends only on $T$ and on a quantity depending on the correlation length of the waves. This can be better seen using an analytic cross-covariance given by a Lorentzian of the form
\begin{equation}
 \Ecc(\xv, \xv, \omega) = \frac{C_0}{1+ (\omega - \omega_0)^2 / \gamma^2}
\end{equation}
where $\gamma$ is the half width at half maximum of the Lorentzian centered at a frequency $\omega_0$. In this case, one can check that the bound in \Eq{error_bound} is equal to $1/(4 \pi^2 \gamma T)$. Therefore the correlations between frequencies should only be neglected when this bound is small, i.e.\ the observation time is long enough to represent correctly the mode.

As the covariance between travel times is known to be also of order $1/T$ \citep{GIZ04}, it is legitimate to wonder if the frequency correlations should be taken into account. It is shown below (see \Eq{covtau}) 
that considering frequency correlations will only lead to additional terms of order $1/T^2$ that can be neglected for long observation times. 

\section{Model noise covariances} \label{sect:noiseTau}

In this section and Appendices \ref{sect:convergenceTauApp}--\ref{sect:covtautauApp}  
we present explicit formulae for the covariance matrices of cross-covariances $C$, travel times $\tau$ and products of cross-covariances or travel-times:
\begin{itemize}
\item $\Cov[\tau_1, \tau_2]$ and $\Cov[C_1, C_2]$ which are linked to the fourth order moment of $\phi_T$,
 \item  $\Cov[\tau_1 \tau_2, \tau_3]$ and 
$\Cov[C_1C_2, C_3]$ which requires the knowledge of the sixth order moment of $\phi_T$ \refereeFirst{and is necessary to compute the moment of order four of $\tau$ and $C$},
 \item $\Cov[\tau_1 \tau_2, \tau_3 \tau_4]$ and $\Cov[C_1C_2, C_3C_4]$ which depend on the eighth order moment of $\phi_T$.
\end{itemize}
For the covariance between two complex random variables $X$ and $Y$ 
we will use the convention
\begin{equation}
 \Cov[X,Y] = \EW{X Y^\ast} - \EW{X}\EW{Y^\ast}.
\end{equation}
In particular, as the mean value of the observables is zero, we have
$\Ecc(\xv_1,\xv_2,\omega) 
= \frac{2\pi}{T} \Cov[\phi_T(\xv_2,\omega), \phi_T(\xv_1,\omega)]. 
$

We will show that all moments of cross-covariance functions depend on $\Ecc$ only. Because the travel time measurement procedure is linear in $C$, the moments of the travel-times can be expressed in terms of $\Ecc$ and of the weight functions $W_i$ (see \Eq{W}).

\subsection{Covariance matrix for $C$ and travel times} 

As a first step, we show in \App{highOrder}  
that the covariance between two cross-correlations is given by
\begin{align}
  \left( \frac{T}{2\pi} \right)^2 &\Cov[C(\xv_1, \xv_2, \omega_1), C(\xv_3, \xv_4, \omega_2)] = \nonumber \\
&
\mathbb{E}[\phi^\ast(\xv_1,\omega_1) \phi(\xv_3,\omega_2)]\; 
\mathbb{E}[\phi(\xv_2,\omega_1) \phi^\ast(\xv_4,\omega_2)]\\
& + \mathbb{E}[\phi^\ast(\xv_1,\omega_1) \phi^\ast(\xv_4,\omega_2)]\; 
\mathbb{E}[\phi(\xv_2,\omega_1) \phi(\xv_3,\omega_2)].\nonumber
\end{align}
For a comparison with and a small correction to the corresponding formula 
in \citet{GIZ04} we refer to \App{convergenceTauApp}. 
The covariance between two travel times is given by
\begin{align}
& \Cov[\tau_1, \tau_2] = \frac{(2\pi)^3}{T} \int_{-\pi/\HT}^{\pi/\HT} d\omega W_{\alpha_1}^\ast(\xv_1, \xv_2, \omega) \times  \nonumber\\
 &\quad \Bigl(W_{\alpha_2}(\xv_3, \xv_4, \omega) \Ecc(\xv_1, \xv_3, \omega) \Ecc(\xv_4, \xv_2, \omega)  \label{eq:covtau} \\
 &\quad+ W_{\alpha_2}^\ast(\xv_3, \xv_4, \omega) \Ecc(\xv_1, \xv_4, \omega) \Ecc(\xv_3, \xv_2, \omega) \Bigr) 
+ \frac{X_2}{T^2} + \Mc{O}\left(\frac{1}{T^{m+1}}\right) \nonumber
\end{align}
where $\Mc{O}\left(1/T^{m+1}\right)$ means that the additional terms decay at least as $1/T^{m+1}$ ($m$ corresponds to the regularity, i.e.\ the number of derivatives of the functions 
$\Ecc$ and $W$). A good agreement between the leading order term in 
this formula and SOHO MDI 
measurements was found by \citet{GIZ04}. An explicit formula for the second order term  $X_2$  
is derived in Appendices \ref{sect:convergenceTauApp} and \ref{sect:lemma}. 
If the observation time $T$ is so small that $X_2/T^2$ cannot be neglected, 
$X_2$ can easily be evaluated numerically. 


%

\subsection{Covariance matrix for products of travel times} \label{sect:noiseTautau}

In this section, we are interested in the covariance matrix for the travel times correlations i.e.\ to evaluate the quantity 
\begin{equation}
 \Cov[\tau_1(\mathbf{x}_1, \mathbf{x}_2) \tau_2(\mathbf{x}_3, \mathbf{x}_4), \tau_3(\mathbf{x}_5, \mathbf{x}_6) \tau_4(\mathbf{x}_7, \mathbf{x}_8)]. \label{eq:covtautaudef}
\end{equation}
 This quantity is the most general we can evaluate for velocity correlations. It will be helpful to derive all the formulae in more specific frameworks. In general this quantity depends on the eight points $\xv_i$ but it is of course possible to look at simpler cases.  For example, we may be interested in the correlations between a East-West (EW) and North-South (NS) travel time as presented in \Fig{example}. This quantity can give us informations about the correlations between the velocities $v_x$ and $v_y$, velocities in the $EW$ and $NS$ directions respectively. 

The formula for the product of cross-covariances is given in \App{covtautauApp} (\Eq{covCCCC}) and will not be discussed in the text where we will focus on products of travel times. In \App{covtautauApp}, we derive the general formula for \Eq{covtautaudef} 
\begin{equation}
 \Cov[\tau_1 \tau_2, \tau_3 \tau_4 ] = \frac{1}{T} Z_1 + \frac{1}{T^2} Z_2 + \frac{1}{T^3} Z_3 + \Mc{O}\left(\frac{1}{T^4}\right) \label{eq:covarianceMatrixText}
\end{equation}
where $Z_1$, $Z_2$ and $Z_3$ are given by \Eqss{covtautau1}{Z2}{Z3text} and will be detailed later after some general remarks on this formula. An important point is that all the terms in $Z_i$ depend only on $\Ecc$ and on the weight functions $W$. Thus, it is possible to estimate directly the noise covariance matrix via this formula instead of performing a large number of Monte-Carlo simulations. This strategy is much more efficient as we will see in \Sec{convergenceN} where we demonstrate the rate of convergence of the stochastic simulations.

The terms on the right hand side of the general formula \Eq{covarianceMatrixText} are of different orders with respect to the observation time. The behaviour of these terms is studied in \Sec{convergenceT}. 

Let us now give the expressions for the different terms $Z_i$ in \Eq{covarianceMatrixText}. The term of order $T^{-1}$ is given by (for details, see \App{covtautauApp}):
\begin{align} 
  \frac{1}{T}Z_1 =& \Et_2 \Bigl( \Et_4 \Cov[\tau_1, \tau_3] + \Et_3 \Cov[\tau_1, \tau_4] \Bigr) \nonumber \\
 &+ \Et_1 \Bigl( \Et_4 \Cov[\tau_2, \tau_3] +  \Et_3 \Cov[\tau_2, \tau_4] \Bigr) \label{eq:covtautau1}
\end{align}
where the covariance between two travel times is given by \Eq{covtau} and 
$\Et_j$ is the expectation value of the travel time $\tau_j$, for example,
\begin{equation}
 \Et_1 = \int_{-\pi/\HT}^{\pi/\HT} d\omega W_1^{\ast}(\xv_1, \xv_2, \omega)
\left( \Ecc(\mathbf{x}_{1}, \mathbf{x}_{2}, \omega) - C^{\textrm{ref}}(\mathbf{x}_{1}, \mathbf{x}_{2}, \omega)\right). \label{eq:delta}
\end{equation}
As $C^{\textrm{ref}}$ and $\Ecc$ are generally close or even equal it is possible that this quantity is close to 0 or even exactly 0. This simplification is discussed in \Sec{simplification}. Note that the time dependence (in $T^{-1}$) in \Eq{covtautau1} is hidden on the right hand side in the covariance betweeen two travel times (cf. \Eq{covtau}). 

The term of order $T^{-2}$ is given by:
\begin{align}
  \frac{1}{T^2} Z_2 =& \Cov[\tau_1, \tau_3] \Cov[\tau_2, \tau_4] + \Cov[\tau_1, \tau_4] \Cov[\tau_2, \tau_3] \nonumber \\
 &- \Et_1 \paren{\Cov[\tau_2, \tau_3 \tau_4] +\Et_3\Cov[\tau_2,\tau_4]+\Et_4\Cov[\tau_2,\tau_3]} \nonumber\\
 &- \Et_2 \paren{\Cov[\tau_1, \tau_3 \tau_4] +\Et_3\Cov[\tau_1,\tau_4]+\Et_4\Cov[\tau_1,\tau_3]} \nonumber\\
 &- \Et_3 \paren{\Cov[\tau_1 \tau_2, \tau_4] +\Et_1\Cov[\tau_2,\tau_4]+\Et_2\Cov[\tau_1,\tau_4]} \nonumber\\
 &- \Et_4 \paren{\Cov[\tau_1 \tau_2, \tau_3] +\Et_1\Cov[\tau_2,\tau_3]+\Et_2\Cov[\tau_1,\tau_3]} \label{eq:Z2}
\end{align}
where the covariance involving three travel times is given in the \App{8thorderFinal} by \Eq{covtau3} and the one between two travel times by \Eq{covtau}. As we will see in \Sec{noiseNum} the first line of this term is dominant in most of the applications.

To write down the term $Z_3$ of order $T^{-3}$ we introduce a function $\Gamma_{\alpha_1,\alpha_2}$ such that 
\begin{equation*}
 \Cov[\tau_1, \tau_2] = \frac{(2\pi)^3}{T} \int_{-\pi/\HT}^{\pi/\HT} d\omega \ \Gamma_{\alpha_1, \alpha_2}(\xv_1, \xv_2, \xv_3, \xv_4, \omega)
+ \mathcal{O}(T^{-2}),
\end{equation*}
i.e.\ according to \Eq{covtau}
\begin{align}
 \Gamma_{\alpha_1, \alpha_2}(\xv_1, \xv_2,& \xv_3, \xv_4, \omega) = \nonumber \\
 W_{\alpha_1}^\ast(\xv_1, \xv_2) \Bigl( &W_{\alpha_2}(\xv_3, \xv_4, \omega) \Ecc(\xv_1, \xv_3, \omega) \Ecc(\xv_4, \xv_2, \omega) \nonumber \\
 &+ W_{\alpha_2}^\ast(\xv_3, \xv_4, \omega) \Ecc(\xv_1, \xv_4, \omega) \Ecc(\xv_3, \xv_2, \omega) \Bigr). \label{eq:gamma}
\end{align}
Then the term of order $T^{-3}$ is given by
\begin{align}
\begin{aligned}
 Z_3 = \frac{(2\pi)^7}{T^3} \sum_{\mu \in \Mc{M}} \int_{-\pi/\HT}^{\pi/\HT} d\omega &\Gamma_{\alpha_1, \alpha_{\mu_1}}(\xv_1, \xv_2, \xv_{\mu_1}, \xv_{\mu_2}, \omega) \times \\
 &\Gamma_{\alpha_{\mu_3}, \alpha_{\mu_5}}(\xv_{\mu_3}, \xv_{\mu_4}, \xv_{\mu_5}, \xv_{\mu_6}, \omega) 
\end{aligned}\label{eq:Z3text}
\end{align}
where $\mu = \{ \mu_1, \mu_2, \cdots, \mu_6 \}$ and the subset $\Mc{M}$ contains all $\mu$ satisfying
\begin{equation}
 \left\{ \begin{array}{llll}
          \mu_1 +1 < \mu_2 \textrm{ if } \mu_1 \textrm{ odd } \\
          \mu_1 < \mu_2 \textrm{ if } \mu_1 \textrm{ even } \\
          \mu_3 < \mu_4 < \mu_5 < \mu_6 \\
          {\mu_1, \ldots \mu_6} \in {3, \ldots, 8}.
         \end{array} \right. \label{eq:subsetM}
\end{equation}
$\Mc{M}$ contains 12 elements, so the term $Z_3$ consists in a sum of 12 terms containing a product of the functions $\Gamma$ defined by \Eq{gamma}.

\subsection{Important special cases} \label{sect:simplification}

\subsubsection{Case $C^\textrm{ref} = \Ecc$} 

As $C^{\textrm{ref}}$ is generally choosen as an average value of the observations, we have $C^{\textrm{ref}} = \Ecc$ or at least $C^{\textrm{ref}} \approx \Ecc$. If there is equality then we can simplify the formula given in the previous section because $\Et = 0$. It follows that the term $Z_1$ is zero as are some elements of $Z_2$. Denoting by $\tilde{Z}_2$ the value of $Z_2$ when $C^{\textrm{ref}} = \Ecc$, we have
\begin{equation}
  \frac{1}{T^2} \tilde{Z}_2  =  \Cov[\tau_1, \tau_3] \Cov[\tau_2, \tau_4] + \Cov[\tau_1, \tau_4] \Cov[\tau_2, \tau_3]. \label{eq:noiseFinal}
\end{equation}
This term is of order $T^{-2}$ as each of the covariance in \Eq{noiseFinal} are of order $T^{-1}$.
The noise covariance matrix is now given by the sum of two terms of order $T^{-2}$ and $T^{-3}$:
\begin{align}
 \Cov[\tau_1 \tau_2, \tau_3 \tau_4] &=
  \frac{1}{T^2} \tilde{Z}_2 + \frac{1}{T^3} Z_3 + \Mc{O}\left(\frac{1}{T^4}\right). \label{eq:covarianceMatrixSimpText}
\end{align}

\subsubsection{Case $C^\textrm{ref} \approx \Ecc$}
Suppose now that we do not have equality but $C^{\textrm{ref}} = (1 + \epsilon) \Ecc$ where $\epsilon$ is a small parameter measuring the difference between the reference cross-covariance and their expectation value. In this case $Z_1$ is of order $\epsilon^2$ and the terms that cancelled out previously in $Z_2$ when $C^{\textrm{ref}} = \Ecc$ are of order $\epsilon$. The numerical tests from \Sec{convergenceCref} will confirm that these terms of order $\epsilon$ and $\epsilon^2$ can be neglected so that \Eq{covarianceMatrixSimpText} can be used even if we just have $C^{\textrm{ref}} \approx \Ecc$. 

\subsubsection{Simplified formula}

We have now defined all the terms involved in \Eq{covarianceMatrixText} to compute the covariance of a product of travel times. As one term is of order $T^{-2}$ and the other one of order $T^{-3}$, it will follow that $Z_2$ will dominate for long observation times. In this case, we have the simplified formula:
\begin{align}
 \Cov[\tau_1 \tau_2, \tau_3 \tau_4] =&  \Cov[\tau_1, \tau_3] \Cov[\tau_2, \tau_4] \nonumber \\
 &+ \Cov[\tau_1, \tau_4] \Cov[\tau_2, \tau_3]. \label{eq:noiseSimp}
\end{align}
In the next section, we will show applications of this formula which will validate the model and the simplified formula. In particular, the numerical tests will tell us that \Eq{noiseSimp} can be used if the observation time is more than roughly a few hours.

\section{Examples and comparisons} \label{sect:noiseNum}

\subsection{SDO/HMI power spectrum for p$_1$ ridge}

\begin{figure*}
\sidecaption
\begin{tabular}{cc}
\includegraphics[trim=15cm 0cm 15cm 0cm, clip=true, width=0.3\linewidth]{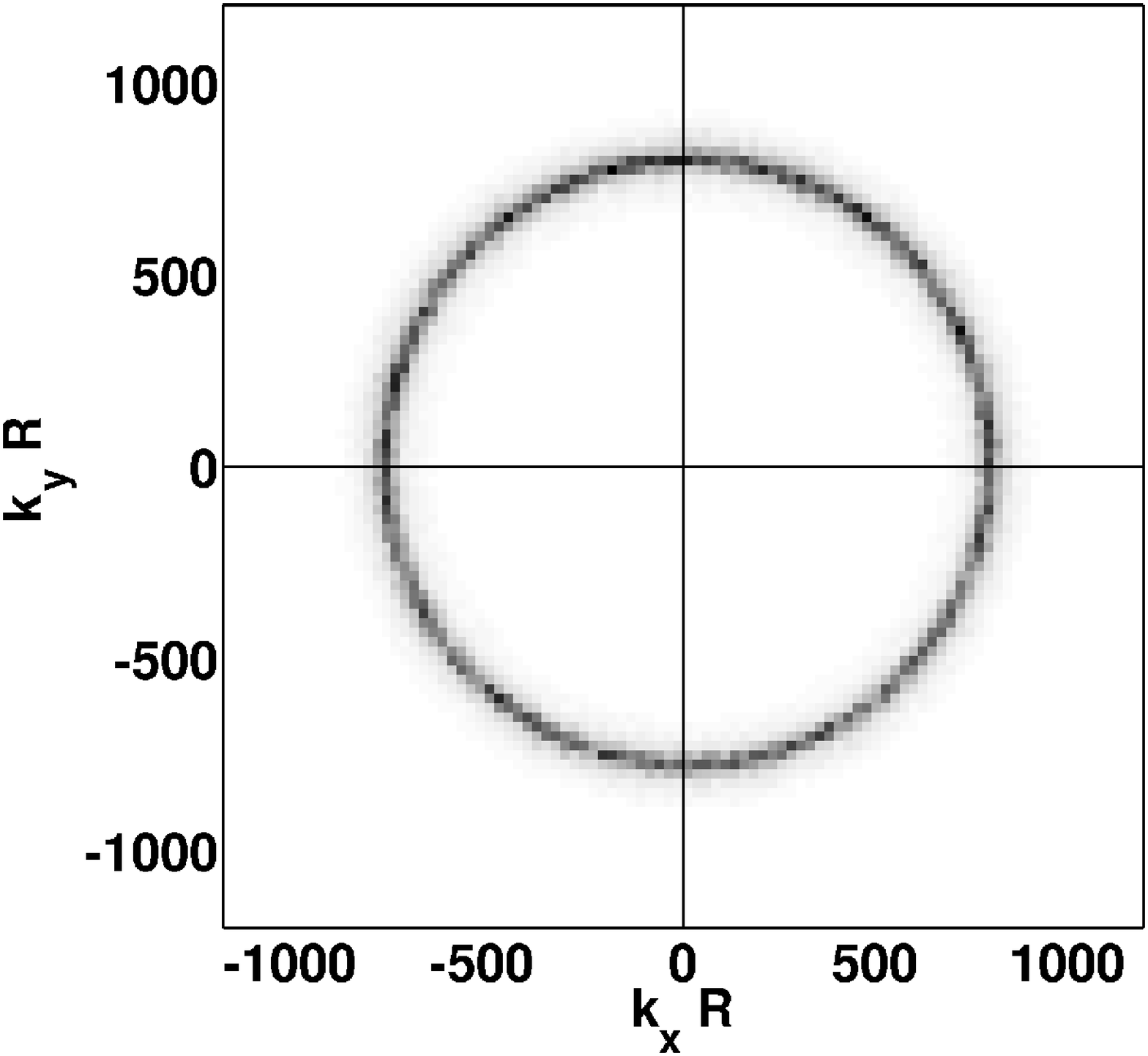} & \includegraphics[trim=15cm 0cm 15cm 0cm, clip=true, width=0.3\linewidth]{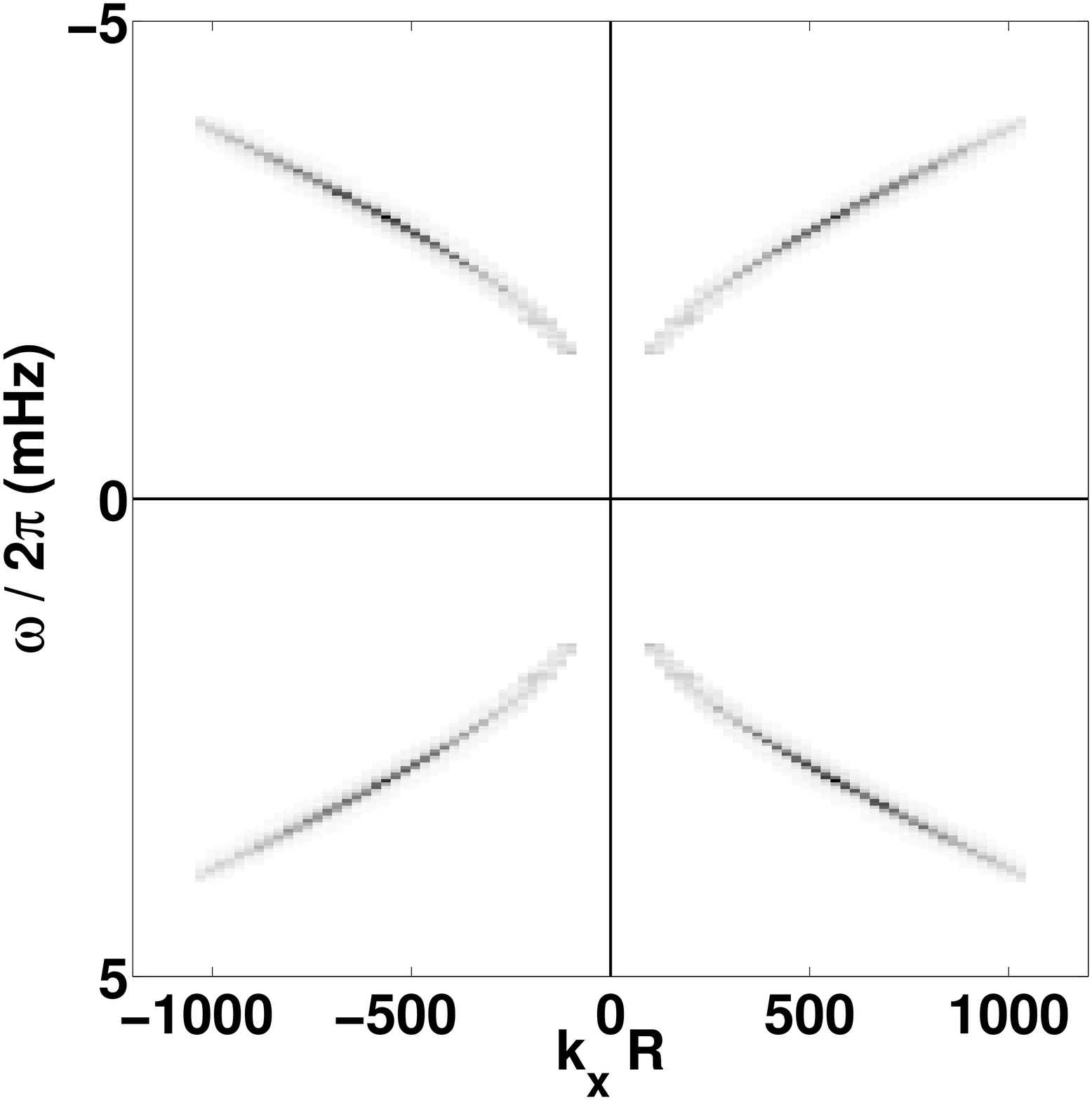}
\end{tabular}
\caption{Average p$_1$ power spectrum ${\Mc P}(\kv,\omega)$ obtained from SDO/HMI dopplergrams.
The sampling is given by $h_k R_\odot = 24.5$ and $h_\omega/2\pi = 34.7$~$\mu$Hz. 
Left panel: cut at frequency $\omega/2\pi=3.4$~mHz. Right panel: cut at $k_y = 0$. The dark parts correspond to large values of the power spectrum, the white one to small values.} \label{fig:power}
\end{figure*}

In this section we validate the analytic formulae for the noise by comparing with Monte Carlo simulations. We choose to use a homogeneous noise so the model depends only on the expectation value of the power spectrum, ${\Mc P}(\kv,\omega) = h_\omega \mathbb{E}[\abs{\phi(\kv,\omega)}^2]$. This expectation value is computed in the Fourier domain in order to perform filtering to keep only the $p_1$ ridge in this case. The quantity ${\Mc P}(\kv,\omega)$ can be estimated from observations by averaging over a set of (quiet-Sun) filtered power spectra $\abs{\phi(\kv,\omega)}^2$. Here we consider observations of line-of-sight Doppler velocity from the HMI instrument on board of the SDO spacecraft \citep{SCH11} between 6 April 2012  and 14 May 2012.
We prepare Postel-projected datacubes of size $N_x \times N_x \times \NT = 512\times512\times610$ centered around the central meridian at a latitude of $40^\circ$. The spatial sampling is $h_x=0.35$~Mm in both directions and the temporal sampling is $\HT = 45$~s. The physical size of the data-cube is $L \times L \times T = 180\, {\rm Mm} \times 180 \,{\rm Mm}  \times 8$~hr. 
The sampling in Fourier space is given by $h_k R_\odot = 24.5$ and $h_\omega/2\pi = 34.7$~$\mu$Hz.

The filtered wave field, $\phi$, is obtained by applying a filter in 3D Fourier space that lets through the p$_1$ ridge only. In this paper we consider only one filter for the sake of simplicity. The function ${\Mc P}(\kv,\omega)$ is estimated by averaging $\abs{\phi(\kv,\omega)}^2$ over forty 8-hr data cubes separated by one day. In \Fig{power}, we show cuts through the average power spectrum.

\subsection{Monte Carlo simulations}
We use the expectation value of the observed power spectrum ${\Mc P}(\kv,\omega)$ defined above as input to the noise model. In order to validate the theoretical model, we run Monte Carlo simulations by generating many realizations of the wave field in Fourier space  using \Eq{noise_model_GB}. The normal distributions are generated with the ziggurat algorithm of MATLAB \citep{MAR84}. All realizations have the same dimensions as above, i.e.\ $h_k R_\odot = 24.5$ and $h_\omega/2\pi = 34.7$~$\mu$Hz.

\subsection{Rate of convergence toward the analytic formula} \label{sect:convergenceN}

To show the importance of having an explicit formula for the noise, we look at the convergence of Monte Carlo simulations to the analytic formula.  For that, we define the following measure of the error:
\begin{equation}
 \textrm{Err}_1(n) = \frac{\abs{\Var[\tau] - \Var_n[\tau]}}{\Var[\tau]}, \label{eq:eps}
\end{equation}
where $\Var[\tau] = \Cov[\tau, \tau]$ is the theoretical variance for travel times computed by \Eq{covtau} and $\Var_n[\tau]$ is the variance obtained by Monte Carlo simulations with $n$ realisations. Similarly, we define
\begin{equation}
 \textrm{Err}_2(n) = \frac{\abs{\Var[\tau^2] - \Var_n[\tau^2]}}{\Var[\tau^2]}, \label{eq:eps2}
\end{equation}
where $\Var[\tau^2] = \Cov[\tau^2, \tau^2]$ is the theoretical variance for a product of travel times computed by \Eq{covarianceMatrixText}.

\begin{figure}
 \includegraphics[width=0.9\linewidth]{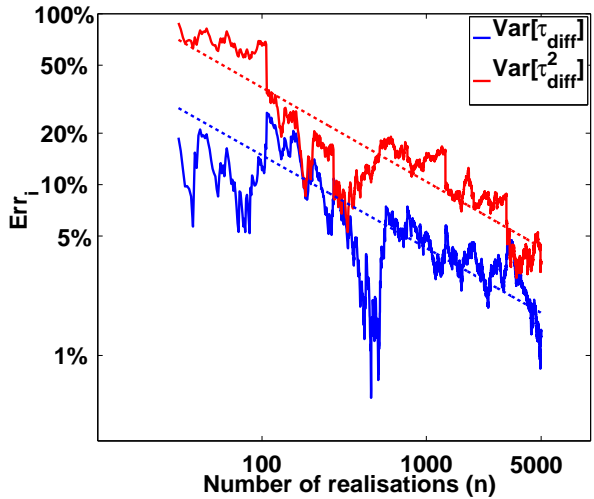}
 \caption{Convergence of the numerical simulations to the model for a $p_1-$ridge with an observation time $T=8$ h. The errors $\textrm{Err}_i(n)$ defined by \Eqs{eps}{eps2} are represented for $\Var[\tau_\textrm{diff}]$ and $\Var[\tau_\textrm{diff}^2]$ for travel times between two points separated by a distance $\Delta = 10$ Mm. The dashed lines has a slope of 1/2 and shows that the error decays as $n^{-\frac{1}{2}}$.} \label{fig:convergenceIni}
\end{figure}
\begin{figure}
\centering
 \includegraphics[width=0.4\linewidth]{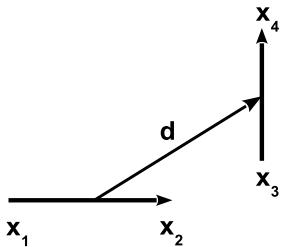}
 \caption{Geometrical configuration $\#$1: geometry used for the covariance between a EW and a NS travel time $\Cov[\tau_1, \tau_2]$ where $\tau_1 = \tau_{\alpha_1}(\xv_1, \xv_2)$ and $\tau_2 = \tau_{\alpha_2}(\xv_3, \xv_4)$. The distance between $\xv_1$ and $\xv_2$ and between $\xv_3$ and $\xv_4$ is $\Delta = 10$ Mm.} \label{fig:example}
\end{figure}

\Fig{convergenceIni} shows the errors $\textrm{Err}_1(n)$  for $\Var[\tau_\textrm{diff}]$ and $\textrm{Err}_2(n)$  for $\Var[\tau_\textrm{diff}^2]$ for travel times between two points separated by a distance $\Delta = 10$Mm. As expected we have
\begin{equation}
 \textrm{Err}_i(n) \approx {\rm const}_i   \,  n^{-\frac{1}{2}}
\end{equation}
with constants depending on the type of measurement. Even if the rate of convergence is the same for $\tau_\textrm{diff}$ or $\tau_\textrm{diff}^2$ the constant is much smaller for a travel time than for a product of travel times. The variance of a product of travel times converges much slower than the travel time variance. For example, an accuracy of $5\%$ is reached with about $n=1000$ realisations for $\tau_\textrm{diff}$ but around $n=5000$ for $\tau_\textrm{diff}^2$.  This underlines the importance of having an analytic formula to obtain the correct limit when $n\rightarrow\infty$, especially in the case of products of travel times.

\subsection{Noise of travel times: comparison with Monte-Carlo simulations and SDO/HMI observations} \label{sect:testTau}

\begin{figure*}
\centering
 \includegraphics[width=0.95\linewidth, trim=4cm 4cm 4cm 4cm]{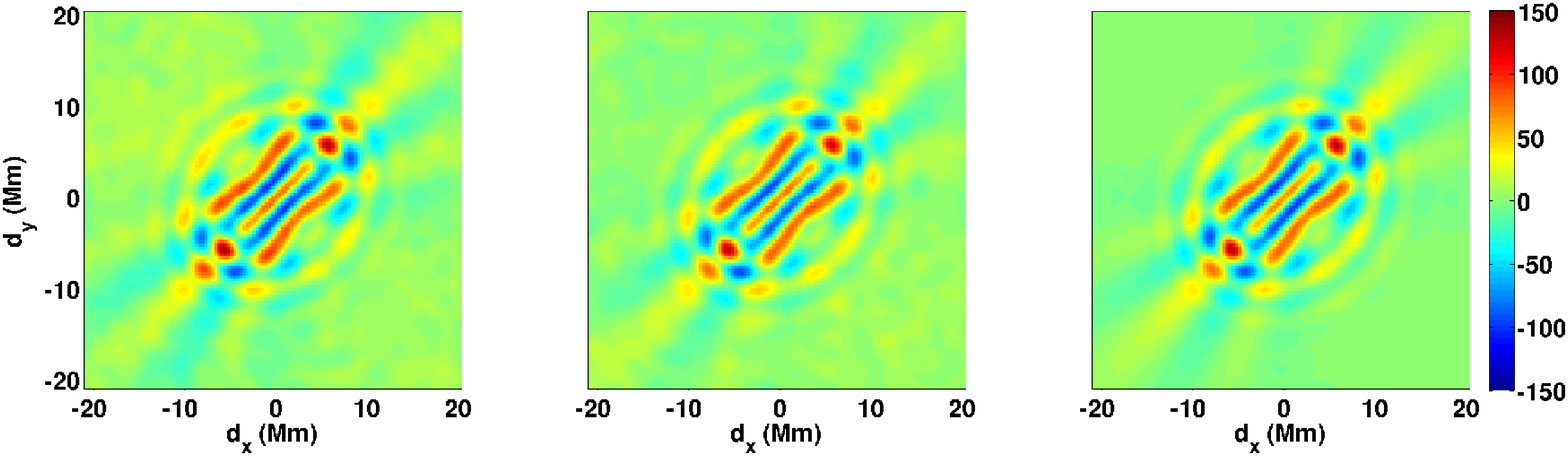} 
 \caption{$\Cov[\tau_1, \tau_2]$ (in s$^2$) for a $p_1-$ridge at a lattitude of $40^\circ$ with an observation time $T = 8$ h in the configuration~$\#$1 given by \Fig{example}. $\tau_+$ is used for $\tau_1$ and $\tau_2$. Left: SDO/HMI observations, middle: Monte Carlo simulation, right: analytic formula } \label{fig:comparisonTau}
\end{figure*}
To show the level of noise in the data, we compare the noise matrix with HMI data from 6 April 2012 until 14 May 2012. The point to point travel times are obtained for a distance $\Delta = 10$Mm in the $x$ and $y$ direction so that we can compare $\Cov[\tau_+(\xv_1, \xv_2), \tau_+(\xv_3, \xv_4)]$ in the configuration given by \Fig{example}. The comparison between the data, Monte Carlo simulation and the explicit formula is given in \Fig{comparisonTau}. As expected, data contain mainly noise as we are looking only at point-to-point travel-times and a good agreement is found between stochastic simulations and the analytic formula.

\subsection{Noise of products of travel times: comparison with Monte-Carlo simulations and SDO/HMI observations} \label{sect:testTautau}

\begin{figure}
\centering
 \includegraphics[width=0.7\linewidth]{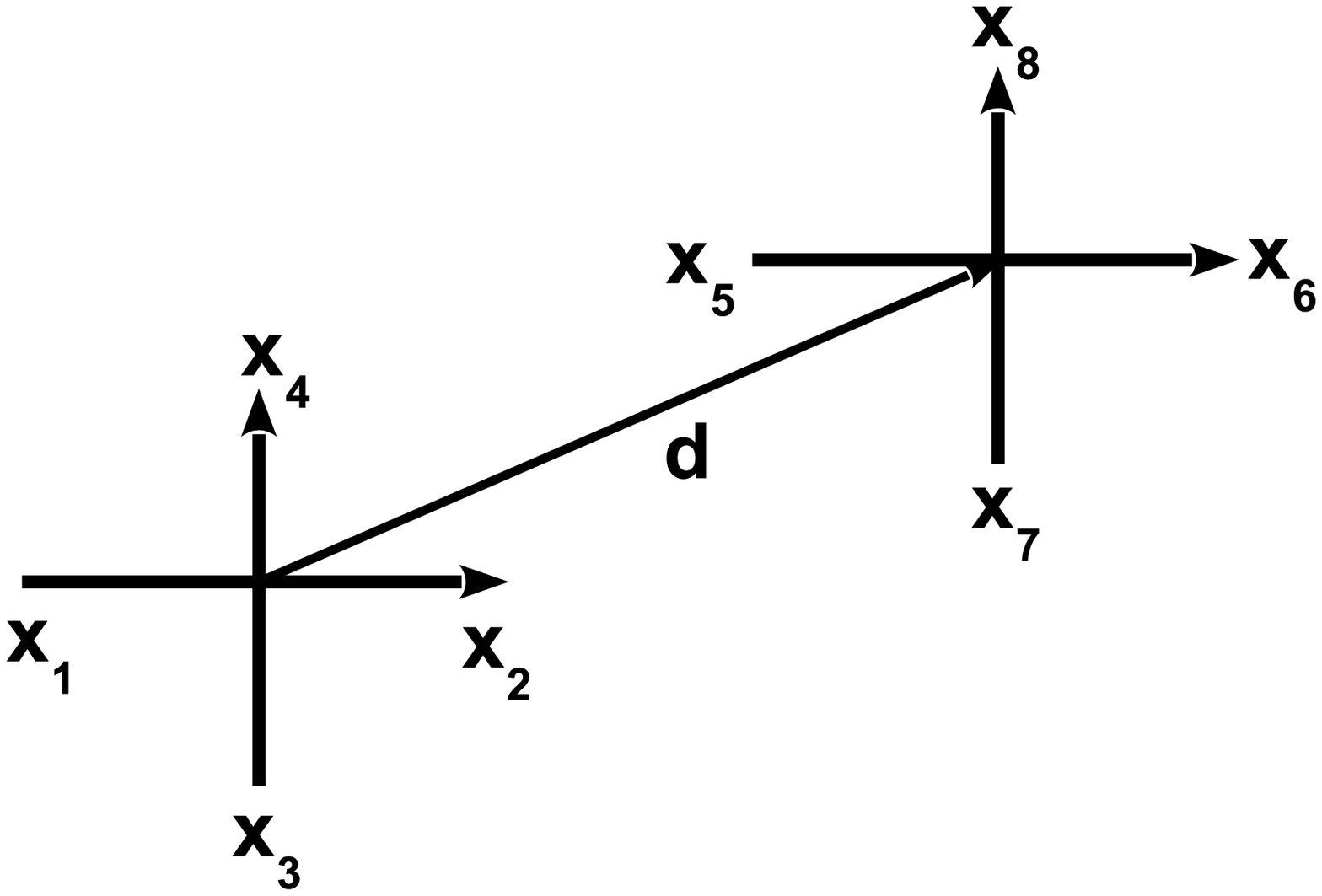}
 \caption{Geometrical configuration~$\#$2: geometry used for the covariance between a product of EW and NS travel times $\Cov[\tau_1 \tau_2, \tau_3 \tau_4]$ where $\tau_i = \tau_{\alpha_i}(\xv_{2i-1}, \xv_{2i})$ are defined in \Eq{taudefinition} . The travel distance between pairs of points is $\Delta = 10$ Mm.} \label{fig:exampleTautau}
\end{figure}

We show in the previous section that the data are dominated by noise in the case of point to point travel times so it is legitimate to ask if there is information in a product of travel times. We look at the covariance between two products of EW and NS travel times $\Cov[\tau_+(\xv_1, \xv_2) \tau_+(\xv_3, \xv_4), \tau_+(\xv_5, \xv_6) \tau_+(\xv_7, \xv_8)]$ as presented in \Fig{exampleTautau}. 

\begin{figure*}
\centering
 \begin{tabular}{ccc}
 \includegraphics[trim=4cm 0cm 3.5cm 0cm, clip=true, width=0.3\linewidth]{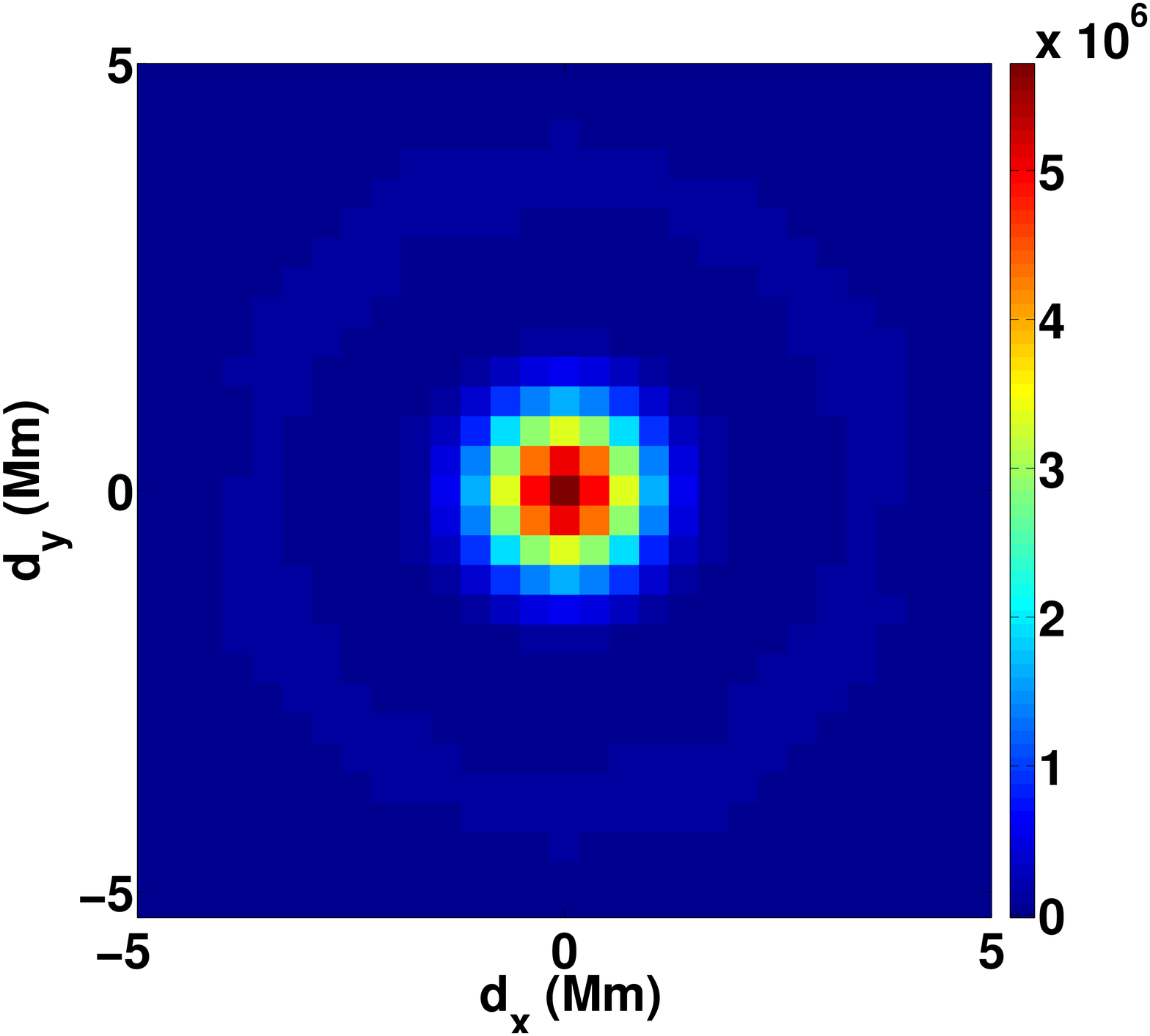} & \includegraphics[trim=4cm 0cm 3.5cm 0cm, clip=true, width=0.3\linewidth]{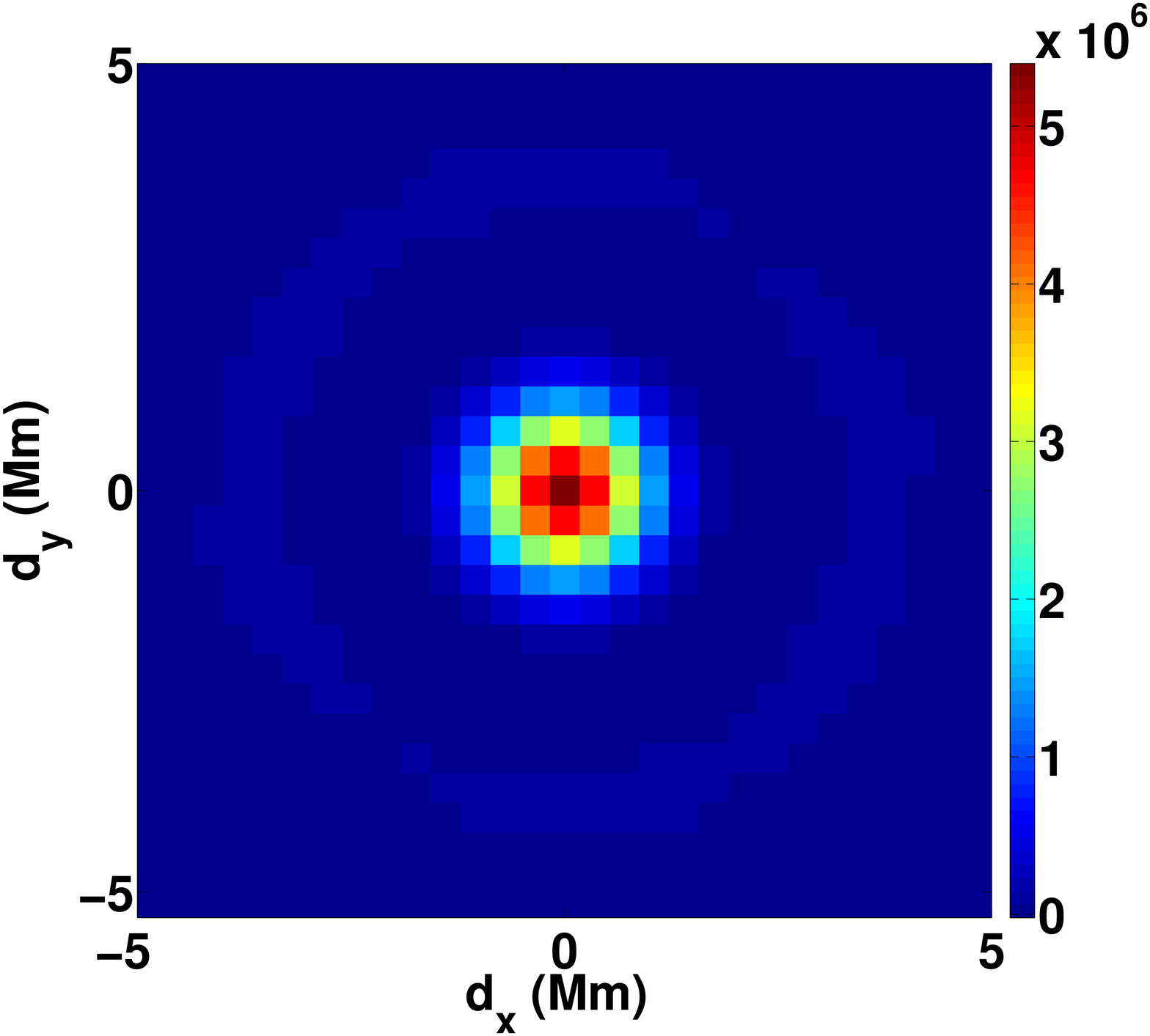} & \includegraphics[trim=4cm 0cm 3.5cm 0cm, clip=true, width=0.3\linewidth]{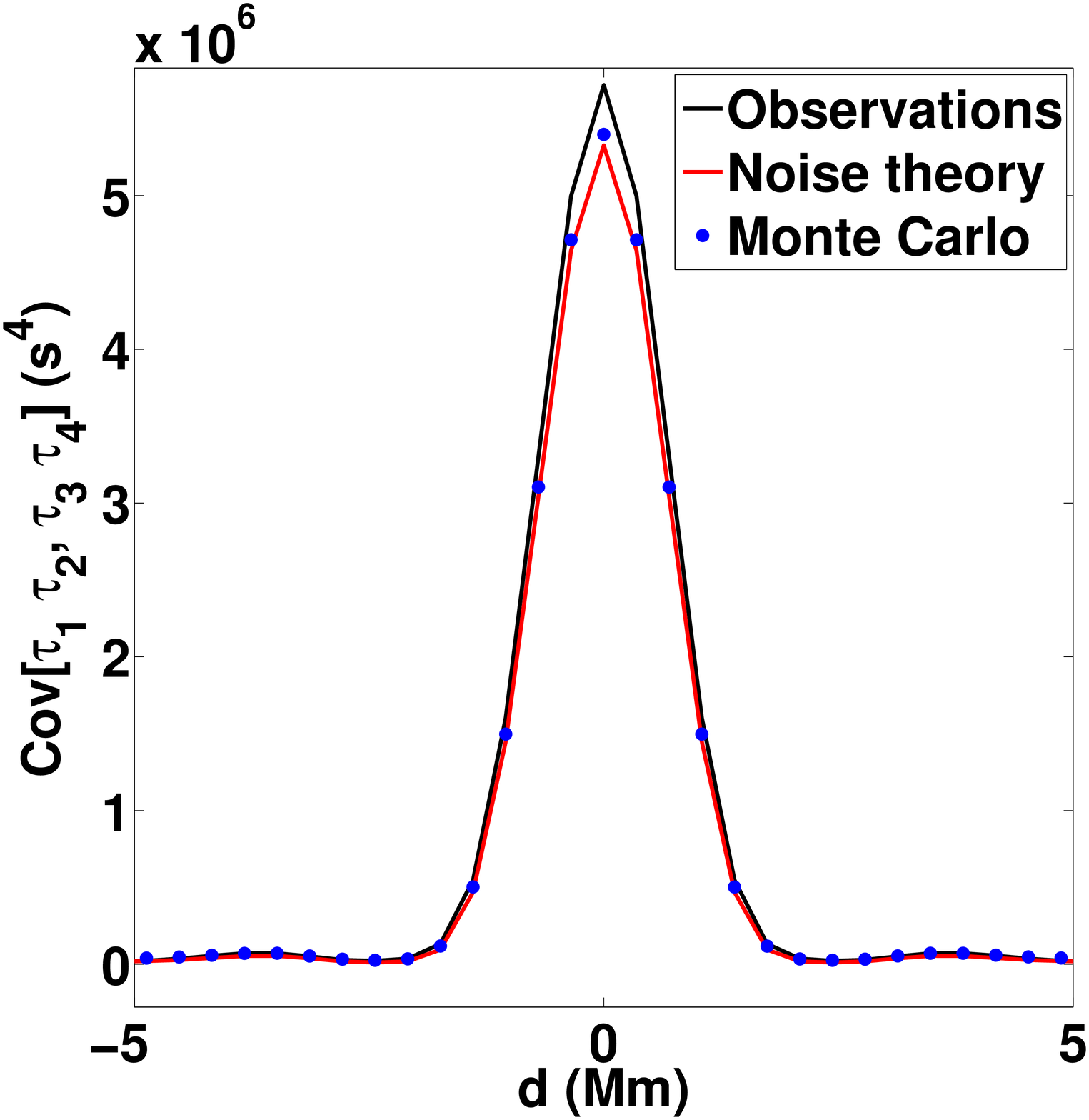}
 \end{tabular}
 \caption{$\Cov[\tau_1 \tau_2, \tau_3 \tau_4]$ (in s$^4$) for a $p_1-$ridge at a lattitude of $40^\circ$ with an observation time $T = 8$ h in the configuration~$\#$2 given by \Fig{exampleTautau}. $\tau_+$ is used for the four travel times. Left: SDO/HMI observations, middle: theory, right: cut through $d_y = 0$ to compare SDO/HMI observations, theory, and Monte Carlo simulations.} \label{fig:comparisonTautau}
\end{figure*}
\begin{figure}
\centering
 \includegraphics[width=0.7\linewidth]{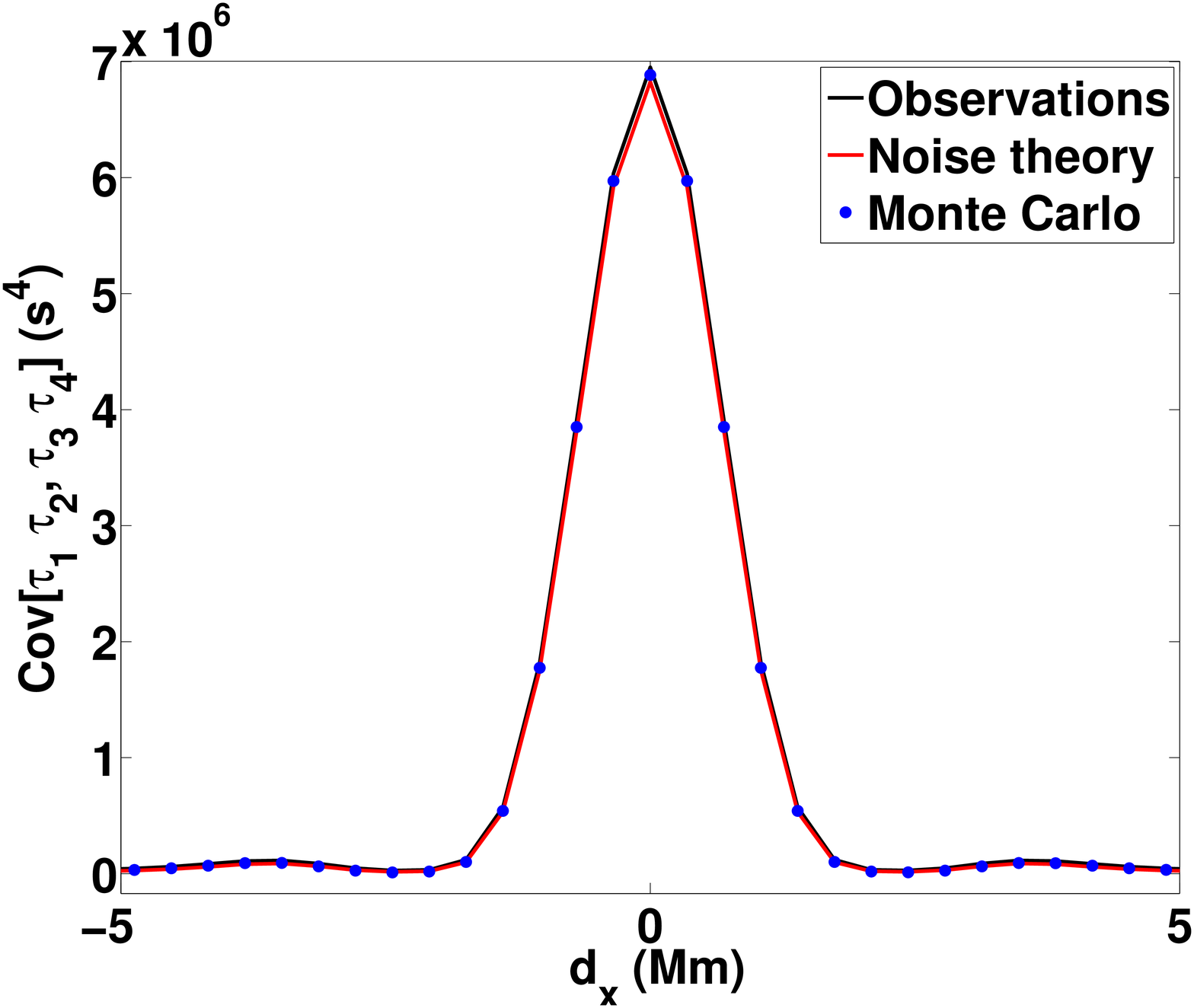}
 \caption{$\Cov[\tau_1 \tau_2, \tau_3 \tau_4]$ (in s$^4$) for a $p_1-$ridge at the equator with an observation time $T = 8$ h in the configuration~$\#$2 given by \Fig{exampleTautau}. This is a cut through $d_y = 0$ comparing SDO/HMI observations, theory, and Monte Carlo simulations.} \label{fig:comparisonTautauEquator}
\end{figure}

The results are given in \Fig{comparisonTautau}. As previously we note a good agreement between the analytic formula and the Monte Carlo simulation. In this case, one can see the differences between the noise and the data \refereeFirst{which are separated by around} $\mathbf{2\sigma}$. To confirm that this difference is due to the presence of physical signal (supergranulation) and not to a problem in the model, we show in \Fig{comparisonTautauEquator} the same covariance but at the equator instead of at a lattitude of $40^\circ$. In this case, data, analytic formula and Monte Carlo simulations fit perfectly. \refereeFirst{Since the product $\langle \tau_x \tau_y \rangle$ (configuration~$\#2$ with $\mathbf{d} = 0$) measures the Reynolds stress $\langle v_x v_y \rangle$, it is expected to be zero at the equator and non-zero away from the equator (as we observe).}

For both lattitudes, the correlation length is identical, equal to $\lambda / 4$ where $\lambda = 7$ Mm is the dominant wavelength of the filtered wave field. \refereeFirst{This is half of the correlation length for travel times as one can see with the simplified formula \Eq{noiseSimp}.}

\subsection{Test of simplified formula for products of travel times using Monte Carlo simulations}

We have shown in \Sec{simplification} that some simplifications can be made to the analytic formula for the noise covariance matrix if $C^\textrm{ref} = \Ecc$. In this section, we show numerically that these simplifications can be done even if we do not have equality and that \Eq{noiseSimp} is a good approximation for the noise covariance matrix.

\subsubsection{Sensitivity to choice of $C^\textrm{ref}$}  \label{sect:convergenceCref}
\begin{figure}
 \centering
 \includegraphics[width=0.7\linewidth, trim=0cm 2cm 0cm 2cm]{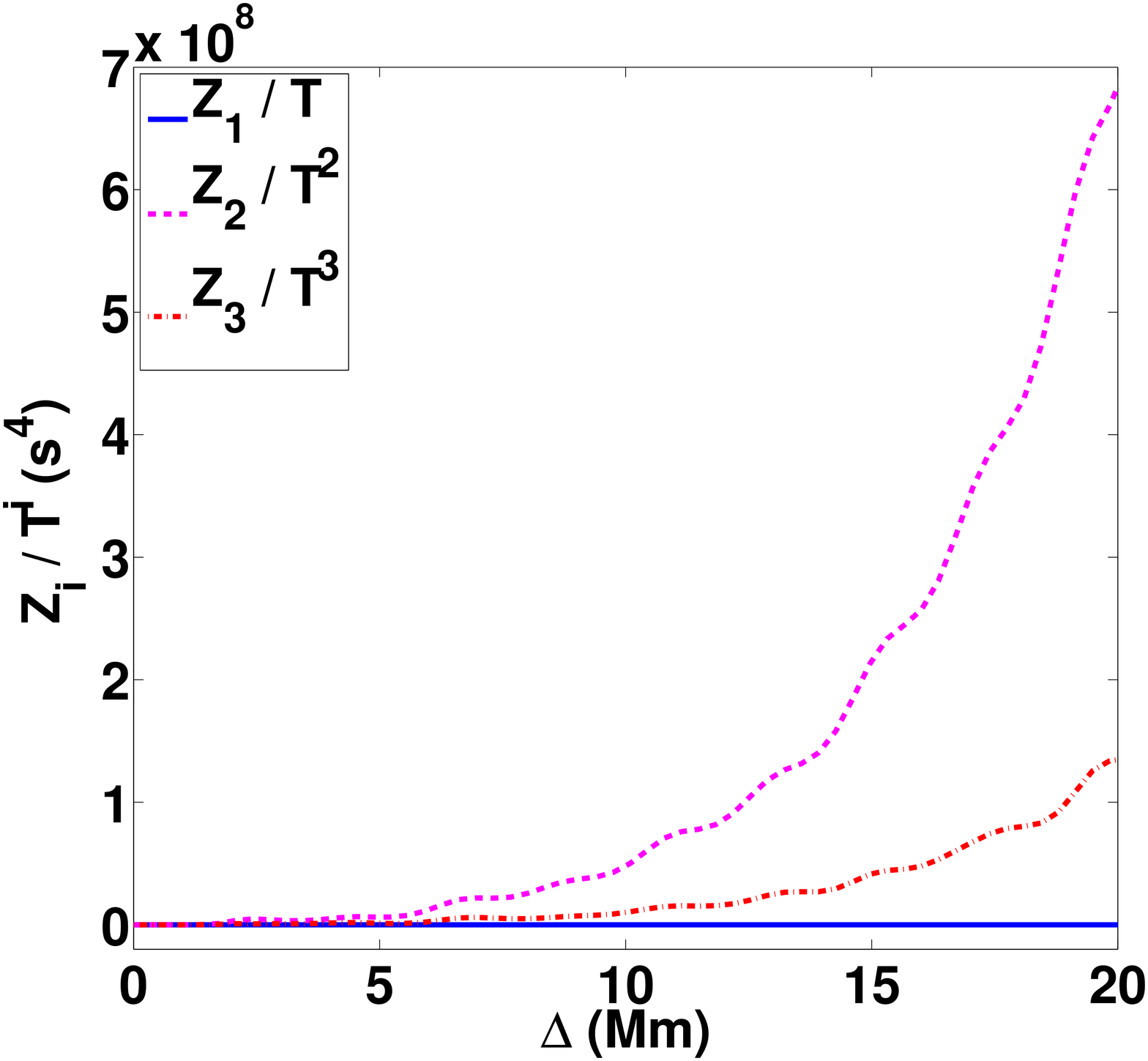}
 \caption{Comparison of the three terms terms in \Eq{covarianceMatrixText} for the variance of a product of travel times separated by a distance $\Delta$. The comparison is done for a $p_1-$ridge and  an observation time of $T = 8$ h.} \label{fig:dependenceCref}
\end{figure}

Let us first consider a fixed observation time ($T = 8$ h for the numerical examples) and look at the dependence on the term $C^\textrm{ref}$. This dependence is due to  the term $Z_1$ and one part of $Z_2$ which depends on $\Et$. \Fig{dependenceCref} makes this comparison for a product of travel times $\tau_\textrm{diff}^2$ between points separated by $\Delta$. In this simple case, it is possible to write down the global behaviour of the different terms in the far field i..e when $\Ecc(\Delta, \omega) < \Ecc(0, \omega)$. If we suppose that $C^{\textrm{ref}} = (1 + \epsilon) \Ecc$ then we have (cf. \App{farFieldApp})
\begin{align*}
 \frac{1}{T}Z_1 & \sim \frac{\epsilon^2}{T} \Ecc(\Delta, \omega)^2 \Ecc(0, \omega)^2  \\
 \frac{1}{T^2}Z_2 & \sim  \frac{1}{T^2} \Ecc(0, \omega)^4 +  \frac{\epsilon}{T^2} \Ecc(0, \omega)^3 \Ecc(\Delta, \omega) \\
 \frac{1}{T^3}Z_3 & \sim \frac{1}{T^3} \Ecc(0, \omega)^4.
\end{align*}
Thus, even if $\epsilon$ is not small, the term $Z_1$ and the second part of the term $Z_2$ are smaller than the other ones in the far field as $\Ecc(\Delta, \omega) < \Ecc(0, \omega)$. This is confirmed in \Fig{dependenceCref} where all the terms are plotted in the worst case, i.e. when $C^\textrm{ref} = 0$. Results are similar for the test cases using the configuration $\#2$ so we did not plot them. Even if the simplifications presented above are only applicable for this particular test case, the terms containing $\Et$ seem to be always smaller than the other ones even when $C^\textrm{ref} = 0$. Thus, as discussed in \Sec{simplification}, when $C^{\textrm{ref}}$ is close to $\Ecc$ and $T$ is not too small it is a good approximation to neglect the terms containing $\Et$ and thus to use \Eq{covarianceMatrixSimpText} to compute the noise covariance matrix.

\subsubsection{Dependence on observation duration $T$} \label{sect:convergenceT}

\begin{figure*}
\centering
\begin{tabular}{cc}
\includegraphics[width=0.45\linewidth]{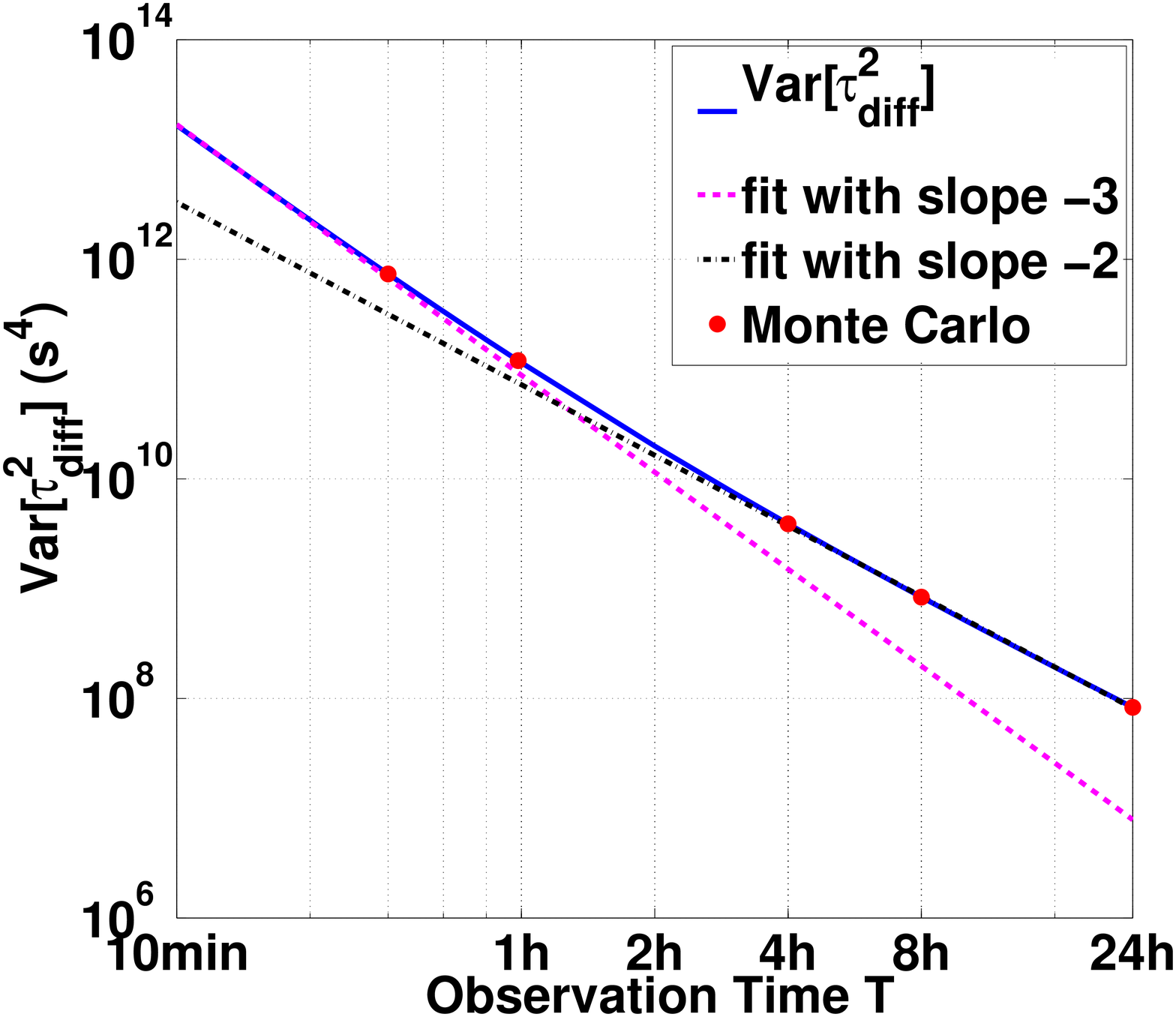} & \includegraphics[width=0.45\linewidth]{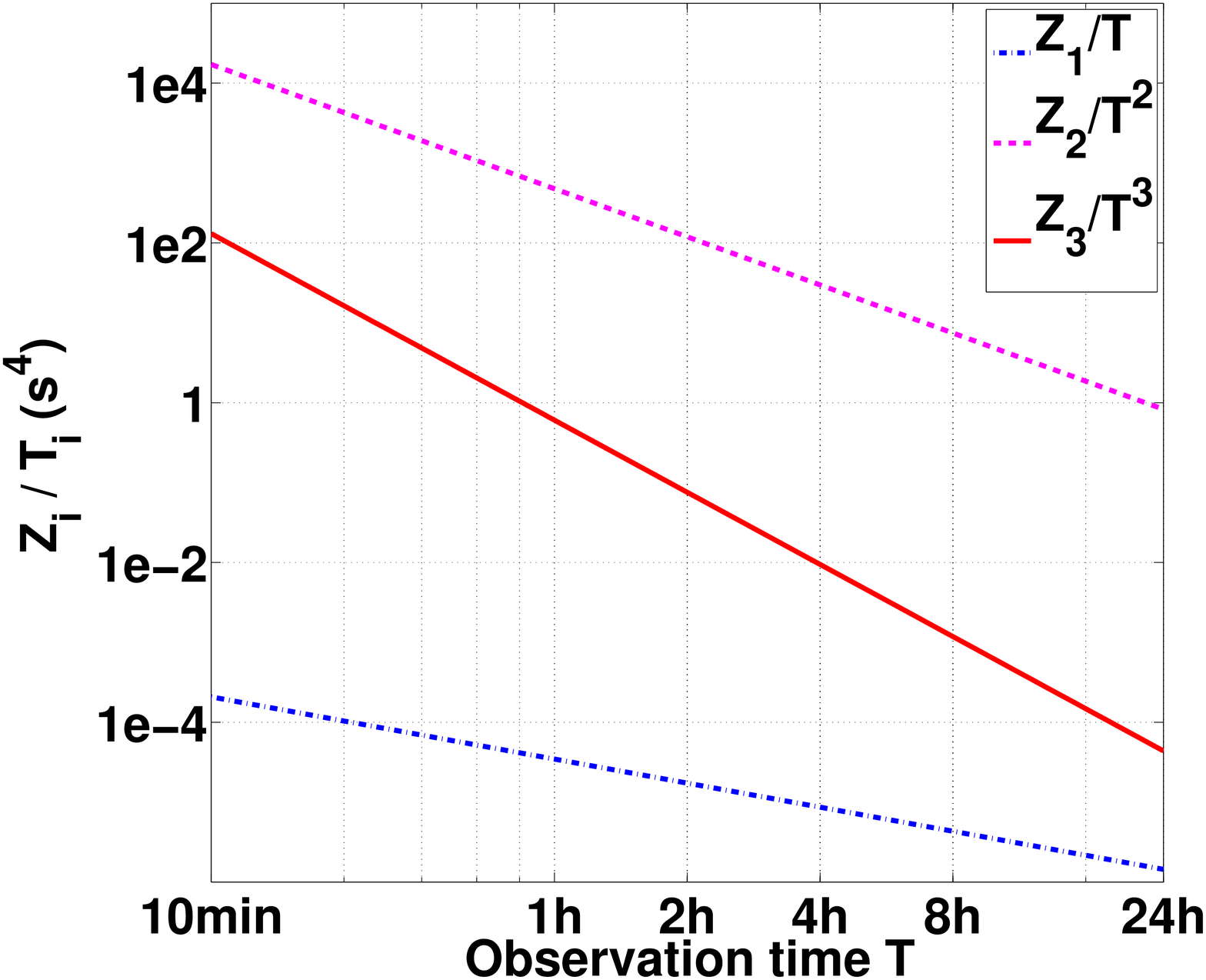}
\end{tabular}
\caption{Left: $\Var[\tau_\textrm{diff}^2]$ as a function of the observation time with $\Delta = 20$ Mm. Right: Comparison of the three terms in \Eq{covarianceMatrixText} for the variance between a product of EW and NS travel time with $\Delta = 20 \textrm{Mm}$. } \label{fig:comparisonTerms}
\end{figure*}

The formula giving the covariance for a product of travel times (\Eq{covarianceMatrixText}) contains three terms that behave differently as a function of the observation time $T$. It is thus interesting to compare these terms to see if some can be dropped or if some are dominant. The term $Z_1$ is initially kept to ensure that the dependence on the observation time will not make this term become significant. As previously, we suppose that we have no knowledge about a reference cross-covariance  ($C^\textrm{ref}$ = 0). \Fig{comparisonTerms} makes this comparison for the variance in the configuration~$\#$1 and the covariance in the configuration~$\#$2 as a function of $T$ (with $\Delta = 20$ Mm).  We see that the contribution of the term $Z_1$ is almost zero so this term can be neglected independently of the observation time. In the first configuration, the term $Z_3$ is always at least two decades smaller than $\tilde{Z}_2$ and so only this last term can be kept. The situation is sligthly 
different for 
the second configuration.
 When $T$ is smaller than one hour, then the standard deviation varies as $T^{-3}$ and the term $Z_3$ is dominant. When the observation time is greater than four hours then it varies as $T^{-2}$ and $\tilde{Z}_2$ is dominant. If $T$ is very long then the variations should be in $T^{-1}$. This area happens theoretically for observation time longer than two monthes which is not realistic for solar applications and is thus not shown in \Fig{comparisonTerms}. The intersection between both terms is given by $T_c = Z_3 / Z_2$. For this test case, a good approximation can be found in the far field as presented in \App{farFieldApp} where it is shown that $T_c \approx 100$min which is confirmed numerically in \Fig{comparisonTerms}. These comparisons of the different terms are extremely important as it implies that we can use the approximation given by \Eq{noiseFinal} if we consider observation times of a few hours which is generally the case. If the observation time is shorter, $\tilde{Z}_2$ is still a  good 
approximation and gives a good estimate of the noise even if the amplitude is not exact. It is certainly sufficient to use $\tilde{Z}_2$ as noise covariance matrix in order to perform an inversion but numerical tests still have to be performed.

\section{Spatial averages} \label{sect:averages}

We define the average value of a quantity $q$ over an area $A$ as follows
\begin{equation}
 \langle q \rangle_{A}   = \frac{1}{A} h_x^2 \sum_{\bx \in A}   q(\bx). \label{eq:tauAverageDef}
\end{equation}
The noise covariance matrix for averaged travel times and products of travel times can be obtained by integrating respectively \Eq{covtau} and \Eq{covarianceMatrixText}. Averaging data has the advantage of increasing the signal-to-noise ratio and allows to deal with fewer data. 
\Tab{average} shows the accuracy of the analytic formula and the importance of the averaging. It compares the value of the variance for a product betweeen EW and NS travel times (configuration $\#2$ with $d=0$) and the same variance when the quantities are averaged over a domain $A = l^2$ with $l = 18$ Mm. First of all, we note a good agreement between the analytic formula and the Monte Carlo simulations. Second, the value of the variance is reduced of a factor 100 when we average the product of travel times over the spatial domain. As expected the variance decreases with the number of independent realisations which is the area $A$ divided by square of the correlation length $\lambda / 4$ (see \Sec{testTautau}) i.e. $18^2 / (7/4)^2 = 105$. Finally, the signal to noise ratio increases with the averaging and we can see a difference due to physical signal between the observations and the noise model.

\begin{table}
\centering
 \begin{tabular}{ccc}
 \hline
  & $\Var(\tau_x \tau_y)$ & $\Var(\langle \tau_x \tau_y \rangle_{A'})$ \\
  & (s$^2$) & (s$^4$) \\
  \hline
  SDO/HMI Observations & $5.7 . 10^6$ & $6.2 . 10^4$\\
  \hline
  Monte Carlo simulations & $5.4 . 10^6$ & $5.0 . 10^4$\\
  \hline
  Analytic formula & $5.4 . 10^6$ & $5.0 . 10^4$\\
  \hline
 \end{tabular}
\caption{$\Var[\tau_1 \tau_2]$ and $\Var[\langle \tau_1 \tau_2 \rangle_{A}]$ (in s$^4$) with $l=18$ Mm for the product of a EW and NS travel time (configuration $\#2$ with $d=0$). Comparison of SDO/HMI observations, analytic formula and Monte Carlo simulations for a $p_1$-ridge at $40^\circ$ lattitude and for an observation time $T=8$ h. 
} \label{tab:average}
\end{table}

\section{Conclusions}
In this paper we presented two main generalizations of the noise model of \citet{GIZ04} for helioseismic travel times. First, the assumption of spatial homogeneity has been dropped. This is useful to model noise in regions of magnetic activity (sunspots and active regions) where oscillation amplitudes are significantly reduced and also to model noise across the solar disk as at different center-to-limb distances. Second, we generalized the noise model to higher-order moments of the travel times, in particular products of travel times. We showed that the covariance matrix for products of travel times consists of three terms that scale like $1/T$, $1/T^2$, $1/T^3$, where $T$ is the total observation time. For standard applications of time-distance helioseismology, we showed that the term in $1/T^2$ is dominant:
\begin{align*}
 \Cov[\tau_1 \tau_2, \tau_3 \tau_4] =& \Cov[\tau_1, \tau_3] \Cov[\tau_2, \tau_4] \\
 &+ \Cov[\tau_1, \tau_4] \Cov[\tau_2, \tau_3].
\end{align*}
This very simple formula links the noise covariance of products of travel times to the covariance of travel times and depends only on the expectation value of the cross-covariance $\Ecc(\xv, \omega)$ and can be obtained directly from the observations. The model is accurate and computationally efficient. It compares very well with Monte Carlo simulations and SDO/HMI observations. The analytic formulae presented in this paper can  be used to compute the noise covariance matrices for averaged quantities and thus increase the signal to noise ratio. Finally we would like to emphasize that our results (moments of order 4, 6, and 8 of the wavefield $\phi(\xv, \omega)$) can be extended to modelling noise for other methods of local helioseismology such as ring-diagram analysis, holography, or far-side imaging.


\begin{acknowledgements}
      The authors acknowledge research funding by Deutsche Forschungsgemeinschaft (DFG) under grant SFB 963/1 ``Astrophysical flow instabilities and turbulence'' (Project A1, ``Solar turbulent convection probed by helioseismology'').
\end{acknowledgements}

\bibliographystyle{bibtex/aa}
\bibliography{biblio}

\onecolumn 
\appendix

\Online
\section{On frequency correlations for the observables} \label{sect:convergenceApp}

In this appendix we study the correlations in frequency space that result 
from a finite observation duration $T$. First we collect some definitions.

Since observations are discrete and to avoid some technical difficulties, 
we only consider discrete time points $t_j=\HT j$, $j\in\Zset$ in this paper. 
As a consequence, the frequency variable $\omega$ 
is $2\pi/\HT$-periodic. However, our definitions of the discrete Fourier 
transform and its inverse are chosen such that we obtain the time-continuous 
case in the limit $\HT\to 0$:
\begin{equation}\label{eq:defFT}
\Mc{P}(\omega)= \frac{\HT}{2\pi}\sum_{k=-\infty}^{\infty}
e^{i\omega t_j}\Mc{P}(t_j),\qquad 
\Mc{P}(t_k) = \int_{-\pi/\HT}^{\pi/\HT} e^{-i\omega t_k}\Mc{P}(\omega)\,
d\omega.
\end{equation}
We will need the orthogonal projection $D_{\NT}$ of $L^2([-\pi/\HT,\pi/\HT])$ onto the space $\Pi_{\NT}$ of 
$2\pi/\HT$-periodic trigonometric polynomials of degree $\leq N$ with the Dirichlet kernel
$\Mc{D}_\NT$, the  Fej\'er smoothing operator 
$F_{\NT}:L^2([-\pi/\HT,\pi/\HT])\to\Pi_N$ with the Fej\'er kernel $\Mc{F}_{\NT}$, and the projected 
periodic Hilbert transform $H_{\NT}:L^2([-\pi/\HT,\pi/\HT])\to\Pi_N$ with kernel $\mathcal{H}_{\NT}$, 
which are defined by
\begin{align*}
 \Mc{D}_{\NT}(\omega) &
= \sum_{k=-\NT}^{\NT} \exp(ik \omega) 
= \begin{cases}\frac{\sin((2\NT+1)\omega/2)}{\sin(\omega/2)},&\omega\neq 0\\
2\NT+1,&\omega=0
\end{cases} &&
(D_{\NT}\Mc{P})(\omega):=\frac{\HT}{2\pi}\int_{-\pi/\HT}^{\pi/\HT}\Mc{D}_{\NT}(\HT(\omega-\tilde{\omega}))
\Mc{P}(\tilde{\omega})\,d\tilde{\omega},\\
 \Mc{F}_{\NT}(\omega) & = \sum_{k=-N}^N \frac{N+1-\abs{k}}{N+1}\exp(ik\omega)
= \begin{cases}\frac{1}{N+1} \frac{\sin^2((\NT+1)\omega/2)}{\sin^2(\omega/2)},&\omega\neq 0\\
\NT+1,&\omega=0
\end{cases} 
&&
(F_{\NT}\Mc{P})(\omega):=\frac{\HT}{2\pi}\int_{-\pi/\HT}^{\pi/\HT}\Mc{F}_{\NT}(\HT(\omega-\tilde{\omega}))
\Mc{P}(\tilde{\omega})\,d\tilde{\omega},\\
 \Mc{H}_{\NT}(\omega) &=\sum_{k=-N}^{N} \frac{\sgn(k)}{i}\exp(ik\omega)
=\begin{cases} \frac{\cos(\omega/2)-\cos((2N+1)\omega/2)}{\sin(\omega/2)},&\omega\neq 0\\
0,&\omega=0
\end{cases}  &&
(H_{\NT}\Mc{P})(\omega):=\frac{\HT}{2\pi}\int_{-\pi/\HT}^{\pi/\HT}\Mc{H}_{\NT}(\HT(\omega-\tilde{\omega}))
\Mc{P}(\tilde{\omega})\,d\tilde{\omega}.
\end{align*}
%
Here $\sgn(k):=1$ and $\sgn(-k):=-1$ for $k\in\Nset$, and $\sgn(0):=0$. 
$H_N$ is related to the standard periodic Hilbert transform $H$ 
with convolution kernel $\Mc{H}(\omega)=\cot(\omega/2)$ by
$H_N=HD_N=D_NH$. 
With our convention for the Fourier transform 
the Fourier convolution theorem is 
$\frac{\HT}{2\pi}\sum_{k=-\infty}^{\infty} f(t_k)\Mc{P}(t_k)e^{i\omega t_k}
= \int_{-\pi/\HT}^{\pi/\HT}f(\omega-\tilde{\omega})\Mc{P}(\tilde{\omega})\,d\tilde{\omega}$. In particular (with 
$f(\omega)=\Mc{F}_{\NT} (\HT\omega)$ and 
$f(t_k)=\frac{2\pi}{\HT}\frac{\NT+1-|k|}{\NT+1}$, etc.) we have
\begin{equation}\label{eq:Opconv}
(D_N\Mc{P})(\omega) = \frac{h_t}{2\pi}\sum_{k=-N}^{N}\Mc{P}(t_k)e^{i\omega t_k},\quad 
(F_N\Mc{P})(\omega) = \frac{h_t}{2\pi}\sum_{k=-N}^{N}
\frac{N+1-|k|}{N+1}\Mc{P}(t_k)e^{i\omega t_k},\quad 
(H_N\Mc{P})(\omega) = \frac{h_t}{2\pi}\sum_{k=-N}^{N}
\frac{\sgn(k)}{i}\Mc{P}(t_k)e^{i\omega t_k}.
\end{equation}

To simplify the notations, the cross-covariance (resp. its expectation value) $C(\xv_a, \xv_b, \omega)$ (resp. $\Ecc(\xv_a, \xv_b, \omega)$) will be simply written as $C_{ab}(\omega)$ (resp. $\Ecc_{ab}(\omega)$) and similarly the weight functions $W(\xv_a, \xv_b, \omega)$ will be $W_{ab}(\omega)$. We will show the following theorem on the correlation function
\begin{equation}
 \Mcc_{ab}(\omega_1, \omega_2) := \frac{2\pi}{T}\mathbb{E}[\phi^\ast(\xv_a, \omega_1) \phi(\xv_b, \omega_2)]: \label{eq:Mcc}
\end{equation}

The covariance between the wavefield at two frequencies $\omega_1$ and $\omega_2$ can be expressed as
\begin{align}
 \Mcc_{ab}(\omega_1, \omega_2) &= 
\begin{cases} I_{ab}(\omega_1, \omega_2) + II_{ab}(\omega_1, \omega_2)& \textrm{for }\omega_1\neq \omega_2,\\
\paren{F_{2\NT} \Ecc_{ab}}(\omega_1)& \textrm{otherwise},
\end{cases}
\label{eq:freq_cor}
\\
\textrm{where } \quad I_{ab}(\omega_1, \omega_2) &:= 
\frac{\HT}{2T} \Mc{D}_{\NT}(\HT (\omega_2 - \omega_1))
\Bigl( \paren{D_{2\NT} \Ecc_{ab}}(\omega_2) + \paren{D_{2\NT}  \Ecc_{ab}}(\omega_1) \Bigr)
\label{eq:covphiTphi_main}\\
 II_{ab}(\omega_1, \omega_2) &:= \frac{\HT}{2T}
\frac{\cos \left( T(\omega_2 - \omega_1)/2 \right)}
{\sin(\HT(\omega_2  -\omega_1) / 2)} 
\Bigl( \paren{H_{2\NT} \Ecc_{ab}}(\omega_1) 
- \paren{H_{2\NT}  \Ecc_{ab}}(\omega_2) \Bigr).
 \label{eq:covphiTphi_pert}
\end{align}
The second term is bounded by
\begin{equation}\label{eq:boundII}
\left| II_{ab}(\omega_1, \omega_2)\right|
\leq \frac{\HT}{4}\left|\sum_{k=-2\NT}^{2\NT}\frac{|t_k|}{T} \Ecc_{ab}(t_k)\right|.
\end{equation}
For stationary Gaussian time series the error of 
the approximate noise model \Eq{GB_model} in \citep{GIZ04} is bounded by 
\begin{equation}
\abs{\Mcc_{ab}(h_{\omega}j,h_{\omega}l) 
- \delta_{j, l} \paren{D_{2\NT}\Ecc_{ab}}\paren{\frac{2\pi j}{T}}} 
\leq \frac{\HT}{4}\left|\sum_{k=-2\NT}^{2\NT}\frac{|t_k|}{T}\Ecc_{ab}(t_k)\right|\qquad \mbox{for}j,l\in\Zset,\; |j|,|l|\leq N. 
\label{eq:result1}
\end{equation}

The proof of the above theorem is given below.

By the definition of $\phi_T$, the covariance betweeen the observations is given by
\begin{align}\label{eq:App_cov1}
\begin{aligned}
\Mcc_{ab}(\omega_1, \omega_2)
&= \frac{\HT^2}{2\pi T}\sum_{l=-\NT}^{\NT} \sum_{k=-\NT}^{\NT} \EW{\phi^\ast(\xv_a, t_l) \phi(\xv_b, t_k)} 
e^{-i\omega_1t_l} e^{i\omega_2t_k}
=  \frac{\HT^2}{2\pi T}\sum_{l=-\NT}^{\NT} \sum_{k=-\NT}^{\NT} \Ecc_{ab}(t_k - t_l) 
e^{-i\omega_1t_l} e^{i\omega_2t_k}\\
 &=  \frac{\HT^2}{2\pi T}\sum_{j=-2\NT}^{2\NT} \Ecc_{ab}(t_j) e^{i\omega_1t_j}  
\sum_{|m|\leq \NT,|j-m|\leq\NT}  e^{-i\omega_1t_{j-m}}e^{i\omega_2t_{j-m}} 
=   \frac{\HT^2}{2\pi T} \sum_{j=-2\NT}^{2\NT} g_{\omega_2 - \omega_1}(j) \Ecc_{ab}(t_j) 
e^{i\omega_1t_j}  
\end{aligned}
\end{align}
where $j=k-l$, $m=-l$, and 
$g_\omega(j) = \sum_{\abs{m}, \abs{j-m} \leq \NT}  e^{i\omega \HT (j-m)}
= \sum_{\abs{m}, \abs{j+m} \leq \NT}  e^{-i\omega \HT m}$. \\ [1ex]
For $\omega_1 = \omega_2$, we have $g_0(j) = 2\NT + 1 - \abs{j}$, so 
\Eq{freq_cor} for this case follows from \Eq{Opconv}. \\[1ex]
 We now consider the case $\omega_1 \neq \omega_2$. For  $j > 0$ we have
\begin{equation*}
g_\omega(j) = \sum_{m=-\NT}^{\NT-j} e^{-i\omega \HT m} 
= e^{i\omega \HT \NT} \frac{1 - e^{-i\omega \HT (2\NT-j+1)}}{1 - e^{-i\omega \HT}} 
= e^{i\omega \HT (\NT+1/2)} \frac{1 - e^{-i\omega \HT (2\NT-j+1)}}
{e^{i\omega\HT/2} - e^{-i\omega \HT/2}} 
= e^{i\omega \HT j/2} \frac{\sin(\HT\omega (2\NT-j+1)/2)}{\sin(\HT \omega / 2)}.
\end{equation*}
If $t_j<0$, then $g_\omega(j) = g^\ast_\omega(-j)$.
Inserting the expression for $g_\omega$ in \Eq{App_cov1}, 
using the identity $\sin(x-y)=\sin x\cos y-\cos x\sin y$ for 
$x= T(\omega_2-\omega_1)/2$ and $y= \HT(\omega_2-\omega_1)|j|/2$, 
and finally using \Eq{Opconv} leads to
\begin{align*}
& \Mcc_{ab}(\omega_1, \omega_2) = 
 \frac{\HT^2}{2\pi T\sin(\HT(\omega_2  -\omega_1) / 2)}  \sum_{j=-2\NT}^{2\NT}  \Ecc_{ab}(t_j) e^{i(\omega_1+\omega_2)t_j/2} \sin \left( \HT \frac{\omega_2 - \omega_1}{2}(2\NT+1-\abs{j}) \right)  \\
 &\qquad= \frac{\HT^2}{4\pi T\sin(\HT(\omega_2  -\omega_1) / 2)} \left( \sin \left( \frac{\omega_2 - \omega_1}{2}T \right)   \sum_{j=-2\NT}^{2\NT}  \Ecc_{ab}(t_j) \left( e^{i\omega_1t_j} + e^{i\omega_2t_j} \right)   
-\cos \left( \frac{\omega_2 - \omega_1}{2}T \right)  \sum_{j=-2\NT}^{2\NT} \Ecc_{ab}(t_j) 
\frac{\sgn(j)}{i} \left( e^{i\omega_2t_j} - e^{i\omega_1t_j} \right) \right)\\
&\qquad=\frac{\HT}{2T} \frac{\sin((\omega_2-\omega_1)T/2)}{\sin(\HT(\omega_2  -\omega_1) / 2)}
\paren{(D_{2\NT}\Ecc_{ab})(\omega_1)+(D_{2\NT}\Ecc_{ab})(\omega_2)}
- \frac{\HT}{2T} \frac{\cos((\omega_2-\omega_1)T/2)}{\sin(\HT(\omega_2  -\omega_1) / 2)}
\paren{(H_{2\NT}\Ecc_{ab})(\omega_2)-(H_{2\NT}\Ecc_{ab})(\omega_1)}.
\end{align*}

To bound $II_{ab}$ we may assume without loss of generality 
that $|\omega_2-\omega_1|\leq \pi/\HT$ due to $2\pi/\HT$-periodicity.
Using the mean value theorem and \Eq{Opconv} and the inequality
$\frac{|x|}{|\sin(x)|}\leq \frac{\pi/2}{\sin(\pi/2)}=\frac{\pi}{2}$ 
for $|x|\leq \frac{\pi}{2}$ we obtain
\begin{align}
\left|\frac{(H_{2\NT}\Ecc_{ab})(\omega_2)-(H_{2\NT}\Ecc_{ab})(\omega_1)}{\sin(\HT(\omega_2-\omega_1)/2)}\right|
&\leq \frac{|\omega_2-\omega_1|}{\sin(\HT(\omega_2-\omega_1)/2)}
\sup_\omega \left|(H_{2\NT}\Ecc_{ab})'(\omega)\right| \nonumber \\
&\leq \frac{1}{\pi}\frac{\HT|\omega_2-\omega_1|/2}{\sin(\HT(\omega_2-\omega_1)/2)}
\left|\sum_{k=-2\NT}^{2\NT}|t_k| \Ecc_{ab}(t_k)\right|
\leq \frac{1}{2}\left|\sum_{k=-2\NT}^{2\NT}|t_k| \Ecc_{ab}(t_k)\right|. \label{eq:boundHf}
\end{align}
This yields \Eq{boundII}. It also implies \Eq{result1} for $j\neq l$ since 
$\Mc{D}_N\left(\frac{2\pi(j-l)}{2N+1}\right)=0$, i.e.\ $I_{ab}(h_{\omega} j,h_{\omega} l)=0$. 
To show \Eq{result1} for $j=l$ we use the bound
\[
\left|\paren{D_{2\NT}\Ecc_{ab}-F_{2\NT}\Ecc_{ab}}(\omega)\right|
\leq \frac{\HT}{2\pi} \left|\sum_{k=-2\NT}^{2\NT} 
\frac{k}{4\NT+2} \Ecc_{ab}(t_k)\right|
= \frac{\HT}{4\pi}\left|\sum_{k=-2\NT}^{2\NT} \frac{|t_k|}{T} 
 \Ecc_{ab}(t_k)\right|
\leq \frac{\HT}{4}\left|\sum_{k=-2\NT}^{2\NT} \frac{|t_k|}{T} 
 \Ecc_{ab}(t_k)\right|.
\]

\section{On frequency correlations for the travel times} \label{sect:convergenceTauApp}
In this appendix we derive the noise covariance matrix for the cross-covariance function $C$ and for the travel time $\tau$ when the frequency correlations are taken into account. \App{convergenceApp} has shown that taking into account the frequency correlations leads to an additional term of order $1/T$ in the covariance of the observables at the grid points. As the covariance betweeen two travel times is also of order $1/T$ it is of interest to look if this correction should be taken into consideration. This appendix proves that the extra term in $1/T$ of the observable covariance will only lead to an additional term in $1/T^2$ for the travel times. We also underline the main difficulties that will occur when computing higher order moments of $C$ and $\tau$.

Since with our convention \Eq{defFT} the Fourier transform is unitary up 
to the factor $\sqrt{2\pi/\HT}$, it follows from definition \eqref{eq:taudef} 
that 
\[
\tau_1(\xv_1, \xv_2) = 2\pi \int_{-\pi/\HT}^{\pi/\HT}
W_{12}^\ast(\omega_1)\left[C_{12}(\omega_1)-
C^{\textrm{ref}}_{12}(\omega_1)\right]\,d\omega_1.
\]
Therefore, 
\begin{equation}
 \Cov[\tau_1(\xv_1, \xv_2), \tau_2(\xv_3, \xv_4)] = (2\pi)^2 \int d\omega_1 \int d\omega_2 \ W_{12}^\ast(\omega_1) W_{34}(\omega_2) \Cov[C_{12}(\omega_1), C_{34}(\omega_2)]. \label{eq:deftautau}
\end{equation}
The first difficulty is to evaluate the quantity $\Cov[C_{12}(\omega_1), C_{34}(\omega_2)]$. For higher order moment we will also need to evaluate
$\Cov [C_{12}(\omega_1) C_{34}(\omega_2), C_{56}(\omega_3)]$ and 
$\Cov [C_{12}(\omega_1) C_{34}(\omega_2), C_{56}(\omega_3) C_{78}(\omega_4)]$.
The way to deal with these terms is presented in \App{highOrder} where it is shown that
\begin{equation}
 \Cov[C_{12}(\omega_1), C_{34}(\omega_2)] = \Mcc_{13}(\omega_1, \omega_2) \Mcc_{42}(\omega_2, \omega_1) + \Mcc_{14}(\omega_1, -\omega_2) \Mcc_{32}(-\omega_2, \omega_1).
\end{equation}
It leads to
\begin{equation}
 \Cov[\tau_1(\xv_1, \xv_2), \tau_2(\xv_3, \xv_4)] = (2\pi)^2  \int d\omega_1 \int d\omega_2 \ W_{12}^\ast(\omega_1) W_{34}(\omega_2) \Bigl( \Mcc_{13}(\omega_1, \omega_2) \Mcc_{42}(\omega_2, \omega_1) + \Mcc_{14}(\omega_1, -\omega_2) \Mcc_{32}(-\omega_2, \omega_1) \Bigr). \label{eq:covtauEx}
\end{equation}
The second difficulty comes from the evaluation of these integrals i.e.\ 
the evalution of linear functionals of the expectation value of the cross-covariance $\Mcc$ given by the weight functions $W$. Similarly, for higher order moments, we will need to be able to evaluate
\begin{align}
 & \int d\omega_1 \int d\omega_2 \int d\omega_3 \  W_{12}(\omega_1) W_{34}(\omega_2) W_{56}(\omega_3) \Mcc_{12}(\omega_1, \omega_2) \Mcc_{34}(\omega_1, \omega_3) \Mcc_{56}(\omega_2, \omega_3) \\
 & \int d\omega_1 \int d\omega_2 \int d\omega_3 \int d\omega_4 \ W_{12}(\omega_1) W_{34}(\omega_2) W_{56}(\omega_3) W_{78}(\omega_4)  \Mcc_{12}(\omega_1, \omega_2) \Mcc_{34}(\omega_2, \omega_3) \Mcc_{56}(\omega_3, \omega_4) \Mcc_{78}(\omega_1, \omega_4).
\end{align}
The method to compute these terms is presented in \App{lemma}. Applying the result for the second order moment presented in \App{lemma2} leads to the result:

The travel-time covariance for finite $T$ is given by the travel-time covariance for infinite observation time (\Eq{covtau}) plus a correction that decreases as $1/T^2$
\begin{align}\label{eq:result2}
\begin{aligned}
 \Cov[\tau_1(\xv_1, \xv_2), \tau_2(\xv_3, \xv_4)] =& \frac{(2\pi)^3}{T} \int d\omega \ W_{12}^\ast(\omega) \Bigl( W_{34}(\omega) \Ecc_{13}(\omega) \Ecc_{42}(\omega) + W_{34}^\ast(\omega) \Ecc_{14}(\omega) \Ecc_{32}(\omega) \Bigr) \\
 &+ \frac{1}{T^2} \Bigl( \Mc{Y}(W_{12}^\ast, W_{34}, \Ecc_{13}, \Ecc_{42}) + \Mc{Y}(W_{12}^\ast, W_{34}^\ast, \Ecc_{14}, \Ecc_{32}) \Bigr)  + \Mc{O}\left( \frac{1}{T^{m+1}} \right), 
\end{aligned}
\end{align}
where $m$ corresponds to the regularity (the number of derivatives) of the functions $\Ecc_{ab}$ and $W_{ab}$,  and
\begin{align}
 \Mc{Y}(W_1,W_2,f,g) =& -(2\pi)^3\int d\omega \
H_{2\NT}(W_{1} W_{2} fg)'(\omega) \nonumber \\
 & + \frac{\pi^2\HT^2}{2} \int d\omega_1 \int d\omega_2 \ W_{1}(\omega_1) W_{2}(\omega_2) \left( \frac{ H_{2\NT} f(\omega_2) - H_{2\NT} f(\omega_1)}{\sin\left(\HT \frac{\omega_2-\omega_1}{2}\right)} \right) \left( \frac{ H_{2\NT} g(\omega_2) - H_{2\NT} g(\omega_1)}{\sin\left(\HT \frac{\omega_2-\omega_1}{2}\right)} \right). \label{eq:Xdef}
\end{align}

\subsection*{Remark concerning the setting of \citet{GIZ04}}
In \citet{GIZ04}, it was supposed that 
\begin{equation}
 \Mcc_{12}(\omega_1, \omega_2) = \delta_{\omega_1, \omega_2} \Ecc(\xv_2 - \xv_1, \omega_1), \label{eq:GBini}
\end{equation}
so the covariance of $C$ is
\begin{equation}
 \Cov[C_{12}(\omega_1), C_{34}( \omega_2)] = \delta_{\omega_1, \omega_2} \Ecc(\xv_3 - \xv_1, \omega_1) \Ecc(\xv_2 - \xv_4, \omega_1) + \delta_{\omega_1, -\omega_2} \Ecc(\xv_4 - \xv_1, \omega_1) \Ecc(\xv_2 - \xv_3, \omega_1). \label{eq:covC}
\end{equation}
Note that \Eq{covC} is exact. It differs slightly from Eq.~(C8) in \citet{GIZ04} which incorrectly contained an additional term. It leads to the covariance between travel times
\begin{equation}
 \Cov[\tau_1, \tau_2] = \frac{(2\pi)^3}{T} \int d\omega \ W_1^\ast(\xv_2-\xv_1, \omega) \Bigl( W_2(\xv_4-\xv_3, \omega) \Ecc(\xv_3 -\xv_1, \omega) \Ecc(\xv_2 -\xv_4, \omega)  + W_2^\ast(\xv_4-\xv_3, \omega) \Ecc(\xv_4 -\xv_1, \omega) \Ecc(\xv_2 -\xv_3, \omega) \Bigr). \label{eq:covtauA}
\end{equation}
Note that \Eq{covtauA} is identical to Eq.~(28) in \citet{GIZ04} as the extra term in the covariance of $C$ was actually neglected by the authors. Taking into account the frequency correlations, \Eq{GBini} is no longer valid and correction terms have to be added to \Eqs{covC}{covtauA}. These correction terms are given in the previous result.

\section{Noise covariance matrix for high order cross-covariances} \label{sect:highOrder}
In this section we present the way to compute the noise covariance matrices for the cross-covariance function $C$ 
\begin{align}
          &\Cov [C_{12}(\omega_1), C_{34}(\omega_2)]  \label{eq:B2} \\
	  &\Cov [C_{12}(\omega_1) C_{34}(\omega_2), C_{56}(\omega_3)] \label{eq:B3} \\
	  &\Cov [C_{12}(\omega_1) C_{34}(\omega_2), C_{56}(\omega_3) C_{78}(\omega_4)]. \label{eq:B4}
\end{align}
Using in \Eqss{B2}{B3}{B4} that the cross-covariance function can be written as a function of the observables
\begin{equation}\label{eq:defC}
 C_{12}(\omega) = \frac{2\pi}{T} \phi^\ast_1(\omega) \phi_2(\omega)
\qquad \mbox{where } \phi_j(\omega) := \phi(\xv_j, \omega)
\end{equation}
we see that the moments of 4, 6 and 8 of the observables have to be computed. In the next section we present a formula to compute high order moment of Gaussian variables. Then, we will apply this formula to compute \Eqss{B2}{B3}{B4}.

\subsection{Expectation value of high-order  products of Gaussian random variables} \label{sect:gaussian}
We have seen that the moments of order 4, 6 and 8 of the observables have to be computed in order to find the noise covariance matrix for cross-covariances and products of cross-covariances. A formula to compute the $(2J)^\textrm{th}-$order moment of a multivariate complex normal distribution with zero-mean can be found in \citet{ISS18}:
\begin{equation}
 \mathbb{E} \left[ \prod_{i=1}^{2J} z_i \right] = \sum_{(\mu, \nu) \in \Mc{M}^J} \prod_{i=1}^J \mathbb{E}\left[ z_{\mu_i} z_{\nu_i} \right], \label{eq:momentOrderJ}
\end{equation}
where $\mu$ and $\nu$ have distinct values in $\llbracket 1, 2J \rrbracket$ and the set $\Mc{M}^J$ is defined by
\begin{equation}
 \Mc{M}^J = \left\{ (\mu_i, \nu_i) \textrm{ with } \mu_i, \nu_i \in \llbracket 1, 2J \rrbracket , \textrm{ s.t. } \mu_i < \nu_i \textrm{ and } (\mu_i)_i \textrm{ increasing} \right\}. \label{eq:subset}
\end{equation}
Here, we used the notation $\llbracket 1, 2J \rrbracket$ for the set of all integers between 1 and $2J$.
In order to better understand \Eq{momentOrderJ} let us explain it for the case $J=2$. In this case, \Eq{momentOrderJ} can be written as
\begin{equation}
  \mathbb{E} [z_1 z_2 z_3 z_4] = \sum_{i,j,k,l} \mathbb{E}[z_i z_j] \mathbb{E} [z_k z_l],
\end{equation}
where the indices $i,j,k,l$ must satisfy $i < j$, $ \ i < k$ and $k < l$ according to \Eq{subset}. This enforces that $i=1$. Then, we can have $k=2$ or $k=3$. If $k=3$, then $l=4$ and $j=2$. If $k=2$ then we have again two possibilities: $l=3$ and so $j=4$ or $l=4$ and $j=3$. So three combinations are possible: $(1,2,3,4)$, $(1,4,2,3)$ and $(1,3,2,4)$. It leads to
\begin{equation}
 \mathbb{E}[z_1 z_2 z_3 z_4] = \mathbb{E}[z_1 z_2] \mathbb{E}[z_3 z_4] + \mathbb{E}[z_1 z_3] \mathbb{E}[z_2 z_4] + \mathbb{E}[z_1 z_4] \mathbb{E}[z_2 z_3]. \label{eq:moment4}
\end{equation}
In particular, we have
\begin{equation}
\Cov(z_1^*z_2,z_3^*z_4) = \mathbb{E}[z_1^*z_2z_3z_4^*]-
\mathbb{E}[z_1^*z_2]\mathbb{E}[z_3z_4^*]
=\mathbb{E}[z_1^*z_3]\mathbb{E}[z_2z_4^*]+
\mathbb{E}[z_1^*z_4^*]\mathbb{E}[z_2z_3],
\label{eq:cov2}
\end{equation}
which is the formula required to compute the moment of order 4 in \Eq{B2}.
For $J=3$, \Eq{momentOrderJ} becomes
\begin{equation}
  \mathbb{E} [z_1 z_2 z_3 z_4 z_5 z_6] = \sum_{i,j,k,l,m,n} \mathbb{E}[z_i z_j] \mathbb{E} [z_k z_l] \mathbb{E} [z_m z_n],
\end{equation}
where the indices $i,j,k,l,m,n$ must satisfy $i < k < m$ (since the sequence $(\mu_i)$ must increase) and $i < j$, $ \ k < l$ and $ \ m < n$ (since $\mu_i < \nu_i$) according to \Eq{subset}. Hence we obtain
\begin{align}\label{eq:order6th}
\begin{aligned}
 \Cov(z_1^* z_2 z_3^* z_4, z_5^* z_6) &= \mathbb{E}[z_1^* z_2 z_3^* z_4 z_5 z_6^{\ast}] - \mathbb{E}[z_1^* z_2 z_3^* z_4] \mathbb{E}[z_5z_6^{\ast}] \\
 &= \mathbb{E}[z_1^* z_2] \mathbb{E}[z_3^* z_5] \mathbb{E}[z_4z_6^{\ast}] +  \mathbb{E}[z_1^*z_2] \mathbb{E}[z_3^*z_6^{\ast}] \mathbb{E}[z_4z_5] 
 +  \mathbb{E}[z_1^*z_3^*] \mathbb{E}[z_2z_5] \mathbb{E}[z_4z_6^{\ast}] +  \mathbb{E}[z_1^*z_3^*] \mathbb{E}[z_2z_6^{\ast}] \mathbb{E}[z_4z_5] \\
 &+  \mathbb{E}[z_1^*z_4] \mathbb{E}[z_2z_5] \mathbb{E}[z_3^*z_6^{\ast}] +  \mathbb{E}[z_1^*z_4] \mathbb{E}[z_2z_6^{\ast}] \mathbb{E}[z_3^*z_5] 
  +  \mathbb{E}[z_1^*z_5] \mathbb{E}[z_2z_3^*] \mathbb{E}[z_4z_6^{\ast}] +  \mathbb{E}[z_1^*z_5] \mathbb{E}[z_2z_4] \mathbb{E}[z_3^*z_6^{\ast}]\\
 &+  \mathbb{E}[z_1^*z_5] \mathbb{E}[z_2z_6^{\ast}] \mathbb{E}[z_3^*z_4] +  \mathbb{E}[z_1^*z_6^{\ast}] \mathbb{E}[z_2z_3^*] \mathbb{E}[z_4z_5] 
 +  \mathbb{E}[z_1^*z_6^{\ast}] \mathbb{E}[z_2z_4] \mathbb{E}[z_3^*z_5] +  \mathbb{E}[z_1^*z_6^{\ast}] \mathbb{E}[z_2z_5] \mathbb{E}[z_3^*z_4]. 
\end{aligned}
\end{align}


A problem is that the cardinality of the set $\Mc{M}^J$ is $(4J)! / [(2J)! 4^J]$ \citep{ISS18} increases exponentially. The sum in \Eq{momentOrderJ} contains 3 terms for $J=2$ and 15 for $J=3$ as shown above. Unfortunately for $J=4$ it leads to 105 terms so it is not convenient to write them down explicitely and we will just list the main guidelines in \Sec{8thorder}.

\subsection{Second order moment of $C$}
In the original paper, the fourth order moment of the observables was guessed 
after looking at all the possible cases in the Fourier domain.
Using the formula \Eq{cov2} and the definitions \Eqs{defC}{Mcc} of 
$C_{ab}$ and $\Mcc_{ab}$ and recalling that $\phi_j^*(\omega)=\phi_j(-\omega)$ 
as $\phi_j(t)$ is real-valued, 
the covariance matrix between two cross-covariances is readily computed 
as follows: 
\begin{align}\label{eq:covCCWithCorr}
\begin{aligned}
 \Cov[C_{12}(\omega_1), C_{34}(\omega_2)] &= \left( \frac{2\pi}{T} \right)^2 
\Cov[ \phi^\ast_1(\omega_1) \phi_2(\omega_1), 
\phi^\ast_3(\omega_2) \phi_4(\omega_2)]\\
&= \left( \frac{2\pi}{T} \right)^2 
\paren{\mathbb{E}[\phi_1^\ast(\omega_1) \phi_3(\omega_2)]\; 
\mathbb{E}[\phi_2(\omega_1) \phi_4^\ast(\omega_2)]
 + \mathbb{E}[\phi_1^\ast(\omega_1) \phi_4^\ast(\omega_2)]\; 
\mathbb{E}[\phi_2(\omega_1) \phi_3(\omega_2)]}\\
&=\Mcc_{13}(\omega_1, \omega_2) \Mcc_{42}(\omega_2, \omega_1) + \Mcc_{14}(\omega_1, -\omega_2) \Mcc_{32}(-\omega_2, \omega_1).
\end{aligned}
\end{align}

\subsection{Third order moment of $C$}
In this section we compute the sixth order moment of the observables defined by \Eq{B3}. After writing the cross-correlations as a function of the observables, we need to compute the moment of order 6 of the observables. This can be done using \Eq{order6th} with $z_1=\phi_1(\omega_1)$, $z_2=\phi_2(\omega_1)$, 
$z_3=\phi_3(\omega_2)$, $z_4=\phi_4(\omega_2)$, $z_5=\phi_5(\omega_3)$, and 
$z_6=\phi_6(\omega_3)$. It will turn out that after integration against 
weight functions the order of the different terms in $1/T$ depends 
on their degree of separability. Therefore, we denote by $\Lambda^3_N$ the 
sum of the terms which can be written as product of at most $N$ functions of 
disjoint subsets of the set of variables $\{\omega_1,\omega_2,\omega_3\}$. Then 
\begin{equation}
 \Cov[C_{12}(\omega_1) C_{34}(\omega_2), C_{56}(\omega_3)] 
= \Lambda^3_1(\omega_1,\omega_2,\omega_3) +  \Lambda^3_2(\omega_1,\omega_2,\omega_3), 
\label{eq:covCCC}
\end{equation}
where
\begin{align}
\begin{aligned}
   \Lambda^3_1  &=  \Bigl( \Mcc_{15}(\omega_1, \omega_3) \Mcc_{32}(\omega_2, \omega_1) \Mcc_{64}(\omega_3, \omega_2) + \Mcc_{14}(\omega_1, \omega_2) \Mcc_{62}(\omega_3, \omega_1) \Mcc_{35}(\omega_2, \omega_3) \Bigr) \\
&+ \Bigl( \Mcc_{15}(\omega_1, -\omega_2) \Mcc_{42}(-\omega_2, \omega_1)\Mcc_{63}(\omega_3, -\omega_2) +
\Mcc_{13}(\omega_1, -\omega_2) \Mcc_{62}(\omega_3, \omega_1)\Mcc_{45}(-\omega_2, \omega_3) \Bigr) \\
&+ \Bigl( \Mcc_{14}(\omega_1, \omega_2) \Mcc_{52}(-\omega_3, \omega_1)\Mcc_{36}(\omega_2, -\omega_3) + 
\Mcc_{16}(\omega_1, -\omega_3) \Mcc_{32}(\omega_2, \omega_1)\Mcc_{54}(-\omega_3, \omega_2) \Bigr)  \\
&+ \Bigl( \Mcc_{13}(\omega_1, -\omega_2) \Mcc_{52}(-\omega_3, \omega_1)\Mcc_{46}(\omega_2, \omega_3) +
\Mcc_{16}(\omega_1, -\omega_3) \Mcc_{42}(-\omega_2, \omega_1)\Mcc_{53}(\omega_3, \omega_2) \Bigr), 
\end{aligned}\label{eq:lambda3_1}
\end{align}
and
\begin{align}
\begin{aligned}
 \Lambda^3_2  =& \Ecc_{34}(\omega_2) \Bigl( \Mcc_{15}(\omega_1, \omega_3) \Mcc_{62}(\omega_3, \omega_1)  + 
 \Mcc_{16}(\omega_1, -\omega_3) \Mcc_{52}(-\omega_3, \omega_1) \Bigr)  \\
&+\Ecc_{12}(\omega_1) \Bigl( \Mcc_{35}(\omega_2, \omega_3)\Mcc_{64}(\omega_3, \omega_2) +  
\Mcc_{36}(\omega_2, -\omega_3)\Mcc_{54}(-\omega_3, \omega_2) \Bigr).  \\
=& \Ecc_{34}(\omega_2) \Cov[C_{12}(\omega_1), C_{56}(\omega_3)] + \Ecc_{12}(\omega_1)  \Cov[C_{34}(\omega_2), C_{56}(\omega_3)] 
\end{aligned}\label{eq:lambda3_2}
\end{align}

\subsection{Fourth order moment of $C$} \label{sect:8thorder}
This section is devoted to the computation of the eigth order moment of the observables defined by \Eq{B4}. Writing the cross-correlations as a function of the observables leads to:
\begin{equation}
  \Cov[C_{12}(\omega_1) C_{34}(\omega_2), C_{56}(\omega_3) C_{78}(\omega_4)] =  \left(\frac{2\pi}{T}\right)^4  \Cov[ \phi^\ast_1 \phi_2 \phi^\ast_3 \phi_4, \phi^\ast_5 \phi_6 \phi^\ast_7 \phi_8].
\end{equation}
Here and in the following we omit the argument $\omega_j$
of the observables $\phi_{2j-1}=\phi_{2j-1}(\omega_j)$ and $\phi_{2j}=\phi_{2j}(\omega_j)$. 
As for the moments of order 4 and 6, we can calculate this expression. 
But as explained in \Sec{gaussian}, the moment of order 8 contains 105 terms, 
so we will not write explicitely all the terms. 
As for the moments of order 6 we arrange the terms as
\begin{equation}
 \Cov[C_{12}(\omega_1) C_{34}(\omega_2), C_{56}(\omega_3) C_{78}(\omega_4)] = 
\paren{\Lambda^4_1 + \Lambda^4_2 + \Lambda^4_3}(\omega_1,\omega_2,\omega_3,\omega_4). \label{eq:covCCCC}
\end{equation}
where $\Lambda^4_N$ is the the sum of all terms which can be written as 
product of at most $N$ functions of 
disjoint subsets of the set of variables $\{\omega_1,\omega_2,\omega_3,\omega_4\}
$.
The three terms $\Lambda^4_N$ will be computed below.

\subsubsection*{Expression for $\Lambda^4_3$}
These terms are the ones from the subset given by \Eq{subset} from which in two expectation values, the observables use the same frequencies, for example $\mathbb{E}[\phi_1^\ast \phi_2] \mathbb{E}[\phi_3^\ast \phi_4]$. It leads to the following formula:
\begin{align}\label{eq:lambda4_3}
\begin{aligned}
\left( \frac{T}{2\pi} \right)^4 \Lambda^4_3 =& \mathbb{E}[\phi_1^\ast \phi_5] \mathbb{E}[\phi_2 \phi_6^\ast] \mathbb{E}[\phi_3^\ast \phi_4] \mathbb{E}[\phi_7 \phi_8^\ast]
+\mathbb{E}[\phi_1^\ast \phi_6^\ast] \mathbb{E}[\phi_2 \phi_5] \mathbb{E}[\phi_3^\ast \phi_4] \mathbb{E}[\phi_7 \phi_8^\ast]
+\mathbb{E}[\phi_1^\ast \phi_7] \mathbb{E}[\phi_2 \phi_8^\ast] \mathbb{E}[\phi_3^\ast \phi_4] \mathbb{E}[\phi_5 \phi_6^\ast]  \\
&+\mathbb{E}[\phi_1^\ast \phi_8^\ast] \mathbb{E}[\phi_2 \phi_7] \mathbb{E}[\phi_3^\ast \phi_4] \mathbb{E}[\phi_5 \phi_6^\ast]
+\mathbb{E}[\phi_1^\ast \phi_2] \mathbb{E}[\phi_3^\ast \phi_5] \mathbb{E}[\phi_4 \phi_6^\ast] \mathbb{E}[\phi_7 \phi_8^\ast]
+\mathbb{E}[\phi_1^\ast \phi_2] \mathbb{E}[\phi_3^\ast \phi_6^\ast] \mathbb{E}[\phi_4 \phi_5] \mathbb{E}[\phi_7 \phi_8^\ast] \\
&+\mathbb{E}[\phi_1^\ast \phi_2] \mathbb{E}[\phi_3^\ast \phi_7] \mathbb{E}[\phi_4 \phi_8^\ast] \mathbb{E}[\phi_5 \phi_6^\ast]
+\mathbb{E}[\phi_1^\ast \phi_2] \mathbb{E}[\phi_3^\ast \phi_8^\ast] \mathbb{E}[\phi_4 \phi_7] \mathbb{E}[\phi_5 \phi_6^\ast].
\end{aligned}
\end{align}
Calculating all the expectation values implies
\begin{align*}
\Lambda^4_3 &=   \Ecc_{34}(\omega_2) \Ecc_{87}(\omega_4) \Bigl( \Mcc_{15}(\omega_1, \omega_3) \Mcc_{62}(\omega_3, \omega_1) + \Mcc_{16}( \omega_1, -\omega_3) \Mcc_{52}(-\omega_3, \omega_1) \Bigr) \nonumber \\
&+  \Ecc_{34}(\omega_2) \Ecc_{65}(\omega_3) \Bigl(  \Mcc_{17}(\omega_1, \omega_4) \Mcc_{82}(\omega_4, \omega_1) +  
  \Mcc_{18}(\omega_1, -\omega_4) \Mcc_{72}(-\omega_4, \omega_1) \Bigr) \nonumber  \\
 &+ \Ecc_{12}( \omega_1) \Ecc_{87}(\omega_4) \Bigl( \Mcc_{35}(\omega_2, \omega_3) \Mcc_{64}(\omega_3, \omega_2)   +    
 \Mcc_{36}(\omega_2, -\omega_3) \Mcc_{54}(-\omega_3, \omega_2) \Bigr) \nonumber \\
 &+  \Ecc_{12}(\omega_1) \Ecc_{65}( \omega_3) \Bigl(  \Mcc_{37}(\omega_2, \omega_4) \Mcc_{84}(\omega_4, \omega_2)   +
  \Mcc_{38}(\omega_2, -\omega_4) \Mcc_{74}(-\omega_4, \omega_2)   \Bigr),
 \end{align*}
which can be written in terms of the covariance between two cross-covariance functions
\begin{align*}
 \Lambda^4_3 &= \Ecc_{87}(\omega_4) \Bigl( \Ecc_{34}(\omega_2) \Cov[C_{12}( \omega_1), C_{56}(\omega_3)] +  \Ecc_{12}(\omega_1) \Cov[C_{34}( \omega_2), C_{56}(\omega_3)] \Bigr)  \\
 &+ \Ecc_{65}(\omega_3) \Bigl( \Ecc_{34}(\omega_2) \Cov[C_{12}(\omega_1), C_{78}(\omega_4)] +  \Ecc_{12}(\omega_1) \Cov[C_{34}(\omega_2), C_{78}(\omega_4)] \Bigr).
\end{align*}

\subsubsection*{Expression for $\Lambda^4_2$}
Two kinds of products in \Eq{momentOrderJ} will lead to terms with only two frequency integrals:
\begin{itemize}
 \item in two expectation values, the constraints on $\omega$ are the same, for example $\mathbb{E}[\phi_1^\ast \phi_4] \mathbb{E}[\phi_2 \phi_3^\ast]$ (they will lead to the first two terms in \Eq{C2})
\item in one expectation value, the observables use the same frequencies, for example $\mathbb{E}[\phi_1^\ast \phi_2]$.
\end{itemize}
Computing all the terms, one can show that
\begin{align} \label{eq:C2}
\begin{aligned}
 \Lambda^4_2 =& \Cov[C_{12}(\omega_1), C_{56}(\omega_3)] \Cov[C_{34}(\omega_2), C_{78}(\omega_4)] + \Cov[C_{12}(\omega_1), C_{78}(\omega_4)] \Cov[C_{34}(\omega_2), C_{56}(\omega_3)] \\
 &+ \Ecc_{12}(\omega_1) \Cov[C_{34}(\omega_2), C_{56}(\omega_3) C_{78}(\omega_4)]
 +\Ecc_{34}(\omega_2) \Cov[C_{12}(\omega_1), C_{56}(\omega_3) C_{78}(\omega_4)] \\
 &+\Ecc_{65}(\omega_3) \Cov[C_{12}( \omega_1) C_{34}(\omega_2), C_{78}(\omega_4)]
  +\Ecc_{87}(\omega_4) \Cov[C_{12}(\omega_1) C_{34}( \omega_2), C_{56}(\omega_3)]. 
\end{aligned}
\end{align}
The terms $\Cov[C, C]$ and $\Cov[C C, C]$ appearing in this expression can be computed using \Eqs{covCCWithCorr}{covCCC}.

\subsubsection*{Expression for $\Lambda^4_1$}
All the other terms will lead to terms that contains only one frequency integral in the covariance of the product of travel times. After reorganizing all the terms, one can show that $\Lambda^4_1$ can be written as
\begin{align*}
 \Lambda^4_1 =& \Bigl( \Mcc_{13}(\omega_1, -\omega_2) \Mcc_{52}(-\omega_3, \omega_1) + \Mcc_{15}(\omega_1, \omega_3) \Mcc_{32}(\omega_2, \omega_1) \Bigr) \Bigl( \Mcc_{74}(-\omega_4, \omega_2) \Mcc_{68}(\omega_3, -\omega_4) + \Mcc_{84}(\omega_4, \omega_2) \Mcc_{67}(\omega_3, \omega_4) \Bigr) \nonumber \\
 &+ \Bigl( \Mcc_{13}(\omega_1, -\omega_2) \Mcc_{62}(\omega_3, \omega_1) + \Mcc_{16}(\omega_1, -\omega_3) \Mcc_{32}(\omega_2, \omega_1) \Bigr) \Bigl( \Mcc_{74}(-\omega_4,\omega_2) \Mcc_{85}(\omega_4, \omega_3) + \Mcc_{84}(\omega_4, \omega_2) \Mcc_{75}(-\omega_4, \omega_3) \Bigr) \nonumber \\
 &+ \Bigl( \Mcc_{13}(\omega_1, -\omega_2) \Mcc_{72}(-\omega_4, \omega_1) + \Mcc_{17}(\omega_1, \omega_4) \Mcc_{32}(\omega_2, \omega_1) \Bigr) \Bigl( \Mcc_{74}(-\omega_4, \omega_2) \Mcc_{68}(\omega_3, -\omega_4) + \Mcc_{84}(\omega_4, \omega_2) \Mcc_{67}(\omega_3, \omega_4) \Bigr) \nonumber \\
 &+ \Bigl( \Mcc_{13}(\omega_1, -\omega_2) \Mcc_{82}(\omega_4, \omega_1) + \Mcc_{18}(\omega_1, -\omega_4) \Mcc_{32}(\omega_2, \omega_1) \Bigr) \Bigl( \Mcc_{54}(-\omega_3, \omega_2) \Mcc_{67}(\omega_3, \omega_4) + \Mcc_{75}(-\omega_4, \omega_3) \Mcc_{64}(\omega_3, \omega_2) \Bigr) \nonumber \\
 &+ \Bigl( \Mcc_{14}(\omega_1, \omega_2) \Mcc_{52}(-\omega_3, \omega_1) + \Mcc_{15}(\omega_1, \omega_3) \Mcc_{42}(-\omega_2, \omega_1) \Bigr) \Bigl( \Mcc_{37}(\omega_2, \omega_4) \Mcc_{68}(\omega_3, -\omega_4) + \Mcc_{38}(\omega_2, -\omega_4) \Mcc_{67}(\omega_3, \omega_4) \Bigr) \nonumber \\
 &+ \Bigl( \Mcc_{14}(\omega_1, \omega_2) \Mcc_{62}(\omega_3, \omega_1) + \Mcc_{16}(\omega_1, -\omega_3) \Mcc_{42}(-\omega_2, \omega_1) \Bigr) \Bigl( \Mcc_{37}(\omega_2, \omega_4) \Mcc_{85}(\omega_4, \omega_3) + \Mcc_{38}(\omega_2, -\omega_4) \Mcc_{75}(-\omega_4, \omega_3) \Bigr) \nonumber \\
 &+ \Bigl( \Mcc_{14}(\omega_1, \omega_2) \Mcc_{72}(-\omega_4, \omega_1) + \Mcc_{17}(\omega_1, \omega_4) \Mcc_{42}(-\omega_2, \omega_1) \Bigr) \Bigl( \Mcc_{35}(\omega_2, \omega_3) \Mcc_{68}(\omega_3, -\omega_4) + \Mcc_{85}(\omega_4, \omega_3) \Mcc_{36}(\omega_2, -\omega_3) \Bigr) \nonumber \\
 &+ \Bigl( \Mcc_{14}(\omega_1, \omega_2) \Mcc_{82}(\omega_4, \omega_1) + \Mcc_{18}(\omega_1, -\omega_4) \Mcc_{42}(-\omega_2, \omega_1) \Bigr) \Bigl( \Mcc_{35}(\omega_2, \omega_3) \Mcc_{67}(\omega_3, \omega_4) + \Mcc_{36}(\omega_2, -\omega_3) \Mcc_{75}(-\omega_4, \omega_3) \Bigr) \nonumber \\
 &+ \Bigl( \Mcc_{15}(\omega_1, \omega_3) \Mcc_{72}(-\omega_4, \omega_1) + \Mcc_{17}(\omega_1, \omega_4) \Mcc_{52}(-\omega_3, \omega_1) \Bigr) \Bigl( \Mcc_{36}(\omega_2, -\omega_3) \Mcc_{84}(\omega_4, \omega_2) + \Mcc_{38}(\omega_2, -\omega_4) \Mcc_{64}(\omega_3, \omega_2) \Bigr) \nonumber \\
 &+ \Bigl( \Mcc_{15}(\omega_1, \omega_3) \Mcc_{82}(\omega_4, \omega_1) + \Mcc_{18}(\omega_1, -\omega_4) \Mcc_{52}(-\omega_3, \omega_1) \Bigr) \Bigl( \Mcc_{36}(\omega_2, -\omega_3) \Mcc_{74}(-\omega_4, \omega_2) + \Mcc_{37}(\omega_2, \omega_4) \Mcc_{64}(\omega_3, \omega_2) \Bigr) \nonumber \\
 &+ \Bigl( \Mcc_{16}(\omega_1, -\omega_3) \Mcc_{72}(-\omega_4, \omega_1) + \Mcc_{17}(\omega_1, \omega_4) \Mcc_{62}(\omega_3, \omega_1) \Bigr) \Bigl( \Mcc_{35}(\omega_2, \omega_3) \Mcc_{84}(\omega_4, \omega_2) + \Mcc_{38}(\omega_2, -\omega_4) \Mcc_{54}(-\omega_3, \omega_2) \Bigr) \nonumber \\
 &+ \Bigl( \Mcc_{16}(\omega_1, -\omega_3) \Mcc_{82}(\omega_4, \omega_1) + \Mcc_{18}(\omega_1, -\omega_4) \Mcc_{62}(\omega_3, \omega_1) \Bigr) \Bigl( \Mcc_{35}(\omega_2, \omega_3) \Mcc_{74}(-\omega_4, \omega_2) + \Mcc_{37}(\omega_2, \omega_4) \Mcc_{54}(-\omega_3, \omega_2) \Bigr).
\end{align*}

\section{Evaluation of separable linear functionals of nonseparable 
products of $\Mcc_{ab}$'s} \label{sect:lemma}
In this section we will derive asymptotic expansions of the terms 
\begin{align}
& \int d\omega_1 \int d\omega_2 \ W_{12}(\omega_1) W_{34}(\omega_2) \Mcc_{12}(\omega_1, \omega_2) \Mcc_{34}(\omega_1, \omega_2) \\
 & \int d\omega_1 \int d\omega_2 \int d\omega_3 \ W_{12}(\omega_1) W_{34}(\omega_2) W_{56}(\omega_3) \Mcc_{12}(\omega_1, \omega_2) \Mcc_{34}(\omega_1, \omega_3) \Mcc_{56}(\omega_2, \omega_3) \\
 & \int d\omega_1 \int d\omega_2 \int d\omega_3 \int d\omega_4 \ W_{12}(\omega_1) W_{34}(\omega_2) W_{56}(\omega_3) W_{78}(\omega_4)  \Mcc_{12}(\omega_1, \omega_2) \Mcc_{34}(\omega_2, \omega_3) \Mcc_{56}(\omega_3, \omega_4) \Mcc_{78}(\omega_1, \omega_4)
\end{align}
in $1/T$ as $T\to\infty$ and explicit formulae for the leading order 
terms. Recall that $\Mcc$  defined in \Eq{Mcc} depends on $T$ although 
this is suppressed in our notation. 

\subsection{Functionals of nonseparable products of 
two $\Mcc_{ab}$ functions}\label{sect:lemma2}
In this subsection we will show that 
\begin{align}
 (2\pi)^2\int d\omega_1 \int d\omega_2 \ W_{1}(\omega_1) W_{2}(\omega_2) \Mcc_{12}(\omega_1, \omega_2)  \Mcc_{34}(\omega_1, \omega_2)  =& \frac{(2\pi)^3}{T} \int d\omega \ W_{1}(\omega) W_{2}(\omega) \Ecc_{12}(\omega) \Ecc_{34}(\omega)  \nonumber \\
 &+ \frac{\Mc{Y}(W_{1}, W_{2}, \Ecc_{12}, \Ecc_{34})}{T^2} + \Mc{O}\Bigl( \frac{1}{T^{m+1}} \Bigr), \label{eq:lemma2}
\end{align}
where $\Mc{Y}$ is defined by \Eq{Xdef} if $\Ecc_{12}$ and $\Ecc_{34}$ 
have $m$ derivatives and  $W_{12}$ and $W_{34}$ have 
$m-1$ derivatives.

Plugging \Eq{freq_cor} into the left hand side of \Eq{lemma2} we arrive 
at a sum $(2\pi)^2(X+2Y+Z)$ involving the following three terms: 
\begin{align}
 X :=&  \int d\omega_1 \int d\omega_2 \ W_{1}(\omega_1) W_{2}(\omega_2)  I_{12}(\omega_1, \omega_2) I_{34}(\omega_1, \omega_2) \label{eq:covtau_term1} \\
 Y :=&  \int d\omega_1 \int d\omega_2 \ W_{1}(\omega_1) W_{2}(\omega_2) I_{12}(\omega_1, \omega_2) II_{34}(\omega_1, \omega_2) \label{eq:covtau_term2} \\
 Z :=& \int d\omega_1 \int d\omega_2 \ W_{1}(\omega_1) W_{2}(\omega_2) II_{12}(\omega_1, \omega_2) II_{34}(\omega_1, \omega_2)  \label{eq:covtau_term3}.
\end{align}
We will repeatedly use the following transformation of variables formula 
for functions $f(\omega_1,\omega_2)$ which are $2\pi/\HT$-periodic in 
both variables:
\begin{equation}
\int_{-\pi/\HT}^{\pi/\HT}d\omega_1 \int_{-\pi/\HT}^{\pi/\HT}d\omega_2 \
f(\omega_1,\omega_2) 
= \int_{-\pi/\HT}^{\pi/\HT}d\tilde{\omega}_1 \int_{-\pi/\HT}^{\pi/\HT}
d\tilde{\omega_2} \
f(\tilde{\omega_1}-\tilde{\omega_2},\tilde{\omega}_1+\tilde{\omega}_2),
\qquad 
\begin{pmatrix}
\tilde{\omega}_1	\\ \tilde{\omega}_2
\end{pmatrix}
 = \frac{1}{2}\begin{pmatrix}
\omega_1 + \omega_2\\ \omega_2 - \omega_1
\end{pmatrix},\qquad 
\begin{pmatrix}
\omega_1	\\ \omega_2
\end{pmatrix}
 = \begin{pmatrix}
\tilde{\omega_1}-\tilde{\omega_2}\\ \tilde{\omega}_1+\tilde{\omega}_2
\end{pmatrix}.
\label{eq:COV_2d}
\end{equation}
Note that even though the Jacobian of this transformation of variables 
is $1/2$, no factor appears since on the right hand side we integrate 
over a domain which can be reassembled to two periodicity cells. 

Using \Eq{COV_2d} and noting that $\mathcal{D}_{N}(\omega)^2 = 
(2N+1)\mathcal{F}_{2N}(\omega)=(T/\HT)\mathcal{F}_{2N}(\omega)$, 
the first term can be written as
\begin{align*}
 X = \frac{\HT}{4T} \int d\tilde{\omega}_1 \int d\tilde{\omega}_2 \ W_{1}(\tilde{\omega}_1 - \tilde{\omega}_2) W_{2}(\tilde{\omega}_1 + \tilde{\omega}_2) \mathcal{F}_{2\NT}(2\HT\tilde{\omega}_2) \Bigl(D_{2\NT} \Ecc_{12}(\tilde{\omega}_1 - \tilde{\omega}_2) &+ D_{2\NT} \Ecc_{12}(\tilde{\omega}_1 + \tilde{\omega}_2) \Bigr) \times \\
 &\Bigl(D_{2\NT} \Ecc_{34}(\tilde{\omega}_1 - \tilde{\omega}_2) + D_{2\NT} \Ecc_{34}(\tilde{\omega}_1 + \tilde{\omega}_2) \Bigr).
\end{align*}
We want interpret the inner product as a convolution with $\mathcal{F}_{2N}$ evaluated at $0$. 
First note that by a change of variables $\int d\tilde{\omega}_2 \ \mathcal{F}_{2N}(2\HT\tilde{\omega}_2)g(\tilde{\omega_2})
= \int d\tilde{\omega}_2 \ \mathcal{F}_{2N}(\HT\tilde{\omega}_2)\frac{1}{2}\bracket{g(\tilde{\omega}_2)+g(\tilde{\omega}_2+\pi/\HT)}$. 
Let $f(\omega_1,\omega_2)$ be $2\pi/\HT$-periodic in both arguments 
and define $\tilde{f}(\tilde{\omega}_1,\tilde{\omega}_2):=
f(\tilde{\omega}_1-\tilde{\omega}_2),\tilde{\omega}_1+\tilde{\omega}_2))$. 
Then 
\[
\tilde{f}\paren{\tilde{\omega}_1,\tilde{\omega}_2+\frac{\pi}{\HT}}
= f\paren{\tilde{\omega}_1-\tilde{\omega}_2-\frac{\pi}{\HT},\tilde{\omega}_1+\tilde{\omega}_2+\frac{\pi}{\HT}}
= f\paren{\tilde{\omega}_1-\tilde{\omega}_2+\frac{\pi}{\HT},\tilde{\omega}_1+\tilde{\omega}_2+\frac{\pi}{\HT}}
= \tilde{f}\paren{\tilde{\omega}_1+\frac{\pi}{\HT},\tilde{\omega}_2}, 
\]
and hence
\begin{equation}
\int d\tilde{\omega}_1\int d\tilde{\omega}_2 \ \mathcal{F}_{2N}(2\tilde{\omega}_2)\tilde{f}(\tilde{\omega}_1, \tilde{\omega}_2)
= \frac{1}{2}\int d\tilde{\omega}_1 \ \bracket{\paren{F_{2N}\tilde{f}}
(\tilde{\omega}_1,0)+ \paren{F_{2N}\tilde{f}}\paren{\tilde{\omega}_1+\frac{\pi}{\HT},0}}
= \int d\tilde{\omega}_1 \ \paren{F_{2N}\tilde{f}}(\tilde{\omega}_1,0)
\end{equation}
where $F_{2N}$ always acts on the second argument. 
As $F_{2N}f = D_{2N}f - \frac{1}{T} H_{2N}f'$, it follows that
\begin{equation*}
 X = \frac{2\pi}{T} \int d\tilde{\omega}_1 \ D_{2N}\paren{W_{1}W_{2} 
(D_{2\NT}\Ecc_{12}) (D_{2\NT}\Ecc_{34})}(\tilde{\omega}_1) 
-\frac{2\pi}{T^2} \int d\tilde{\omega}_1 \ 
H_{2N}\paren{W_{1}W_{2}(D_{2N}\Ecc_{12})(D_{2N}\Ecc_{34})}'(\tilde{\omega}_1).
\end{equation*}
Since $|D_{2\NT}\Ecc_{ab} - \Ecc_{ab}|=\mathcal{O}(T^{-m})$, we get 
an additional $\mathcal{O}(T^{-m})$ if we omit the orthogonal 
projections $D_{2\NT}$ in the last equation. 

To bound Y (\Eq{covtau_term2}), we again apply the change of 
variables in \Eq{COV_2d} to obtain
\begin{equation*}
 Y = \frac{\HT^2}{4T^2} \int d\tilde{\omega}_1 \int d\tilde{\omega}_2 \ 
\sin \left( \tilde{\omega}_2T \right) \cos \left( \tilde{\omega}_2T \right) f(\tilde{\omega}_1, \tilde{\omega}_2)
= \frac{\HT^2}{8T^2} \int d\tilde{\omega}_1 \int d\tilde{\omega}_2 \ 
\sin \left( 2\tilde{\omega}_2T \right) f(\tilde{\omega}_1, \tilde{\omega}_2)
\end{equation*}
where $f$ has uniformly bounded derivatives of order $m-1$. 
When $T$ tends to infinity this corresponds to a high order Fourier coefficient and thus can be made as small as desired. In particular, 
by repeated partial integration 
\begin{equation}
 \abs{\int \ f(\tilde{\omega}_1,\tilde{\omega}_2) \sin(2\tilde{\omega}_2 T) 
d\tilde{\omega}_2} \leq \frac{1}{(2T)^{m-1}} \int d\tilde{\omega}_2 \ 
\abs{
\frac{\partial^{m-1} f}{\partial \tilde{\omega_2}^{m-1}}(\tilde{\omega}_1.\tilde{\omega}_2)}. \label{eq:boundFourier}
\end{equation}

The term $Z$ (\Eq{covtau_term3}) can be transformed on the same way and after using that $\cos^2(\tilde{\omega}_2 T) = (1 - \cos(2\tilde{\omega}_2 T))/2$, we find that
\begin{equation*}
 Z = \frac{\HT^2}{8T^2} \int d\omega_1 \int d\omega_2 \ W_{1}(\omega_1) W_{2}(\omega_2) \left( \frac{ H_{2\NT} \Ecc_{12}(\omega_2) - H_{2\NT} \Ecc_{12}(\omega_1)}{\sin\left(\HT \frac{\omega_2-\omega_1}{2}\right)} \right) \left( \frac{ H_{2\NT} \Ecc_{34}(\omega_2) - H_{2\NT} \Ecc_{34}(\omega_1)}{\sin\left(\HT \frac{\omega_2-\omega_1}{2}\right)} \right)
+\mathcal{O}\paren{\frac{1}{T^{m+1}}}
\end{equation*}
where the higher order term comes from $\cos(2\tilde{\omega}_2 T)$ 
in analoy to \Eq{boundFourier}. As $\lim_{n \rightarrow \infty} H_{2N}f = Hf$ and all the terms in the integrals are bounded it follows that $X$ is of order $1/T^2$.
Gathering the expressions for the three terms $X$, $Y$, $Z$ leads to \Eq{lemma2}.

\subsection{Functionals of nonseparable products of 
three $\Mcc_{ab}$ functions}\label{sect:lemma3}
Let $\Mcc$ be defined by \Eq{Mcc} and $W_i$ representing some functions of $\omega$. Then, we have the following expension:
\begin{align}\label{eq:lemma3}
\begin{aligned}
 (2\pi)^3\int d\omega_1 \int d\omega_2 \int d\omega_3 \ W_{1}(\omega_1) W_{2}(\omega_2) W_{3}(\omega_3) & \Mcc_{12}(\omega_1, \omega_2)  \Mcc_{34}(\omega_1, \omega_3)  \Mcc_{56}(\omega_2, \omega_3) =  \\
 &\frac{(2\pi)^5}{T^2} \int d\omega \ W_{1}(\omega) W_{2}(\omega) W_{3}(\omega) \Ecc_{12}(\omega) \Ecc_{34}(\omega) \Ecc_{56}(\omega) + \Mc{O}\Bigl( \frac{1}{T^3} \Bigr). 
\end{aligned}
\end{align}

Using \Eq{freq_cor} in the left hand side of \Eq{lemma3}, four different types of terms have to be studied
\begin{align}
 X :=&  \int d\omega_1 \int d\omega_2 \int d\omega_3 \ W_{1}(\omega_1) W_{2}(\omega_2) W_{3}(\omega_3) I_{12}(\omega_1, \omega_2) I_{34}(\omega_1, \omega_3)  I_{56}(\omega_2, \omega_3) \label{eq:covtau_term1_3} \\
 Y_1 :=&   \int d\omega_1 \int d\omega_2 \int d\omega_3 \ W_{1}(\omega_1) W_{2}(\omega_2) W_{3}(\omega_3) I_{12}(\omega_1, \omega_2) II_{34}(\omega_1, \omega_3)  II_{56}(\omega_2, \omega_3) \label{eq:covtau_term2_3} \\
 Y_2 :=&  \int d\omega_1 \int d\omega_2 \int d\omega_3 \ W_{1}(\omega_1) W_{2}(\omega_2) W_{3}(\omega_3) I_{12}(\omega_1, \omega_2) I_{34}(\omega_1, \omega_3)  II_{56}(\omega_2, \omega_3) \label{eq:covtau_term3_3} \\
 Z :=&   \int d\omega_1 \int d\omega_2 \int d\omega_3 \ W_{1}(\omega_1) W_{2}(\omega_2) W_{3}(\omega_3) II_{12}(\omega_1, \omega_2) II_{34}(\omega_1, \omega_3)  II_{56}(\omega_2, \omega_3)  \label{eq:covtau_term4_3}
\end{align}
where the expressions $I$ and $II$ are given respectively by \Eqs{covphiTphi_main}{covphiTphi_pert}.

We will use the change of variables
\begin{equation}\label{eq:COV_3d}
\int_{Q}d\omega \ f(\omega) = \int_Q \ d\tilde{\omega}f(\omega(\tilde{\omega})),\qquad 
\tilde{\omega} = \frac{1}{3}\left(\begin{smallmatrix}1 &1& 1\\ -1& 1& 0\\ -1& 0& 1\end{smallmatrix}\right)\omega,
\qquad
\omega = \left(\begin{smallmatrix}1 &-1& -1\\  1& 2& -1\\ 1 &-1& 2\end{smallmatrix}\right)\tilde{\omega},\qquad
Q:=[-\pi/\HT,\pi/\HT]^3
\end{equation}
where the Jacobian $1/3$ does not appear for the same reason as in \Eq{COV_2d}. Applying this to $X$ 
we obtain
\begin{align*}
 X = \left( \frac{\HT}{2T} \right)^3 \int d\tilde{\omega}_1 & \int d\tilde{\omega}_2 \int d\tilde{\omega}_3 \ W_{1}(\omega_1) W_{2}(\omega_2) W_{3}(\omega_3) \Mc{D}_{2N}(3\HT \tilde{\omega}_2) \Mc{D}_{2N}(3\HT \tilde{\omega}_3) \Mc{D}_{2N}(3\HT (\tilde{\omega}_3 - \tilde{\omega}_2)) \times \\
 &\Bigl(D_{2\NT} \Ecc_{12}(\omega_1) + D_{2\NT} \Ecc_{12}(\omega_2) \Bigr) \Bigl(D_{2\NT} \Ecc_{34}(\omega_1) + D_{2\NT} \Ecc_{34}(\omega_3) \Bigr) \Bigl(D_{2\NT} \Ecc_{56}(\omega_2) + D_{2\NT} \Ecc_{56}(\omega_3) \Bigr)
\end{align*}
where $\omega_i$ can be replaced by the corresponding value in $\tilde{\omega}_i$. The role of the Fej{\'e}r kernel is played 
by the function 
\begin{align*}
\Mc{F}_{2\NT}^{2D}(\HT\tilde{\omega}_2,\HT\tilde{\omega}_3) &= 
 \frac{\HT}{T} \Mc{D}_{2N}(\HT \tilde{\omega}_2) \Mc{D}_{2N}(\HT \tilde{\omega}_3) \Mc{D}_{2N}(\HT (\tilde{\omega}_3 - \tilde{\omega}_2)) 
= \frac{\HT}{T} \sum_{j,k,l=-N}^N \exp \Bigl( i\HT \bigl( \tilde{\omega}_2 (j-l) + \tilde{\omega}_3 (k+l) \bigr) \Bigr) \\
&= \frac{\HT}{T} \sum_{\abs{m}+\abs{n} \leq 2\NT} \ \  \sum_{o: |o|\leq N, |m+o|\leq N, |n-o|\leq N}
 \exp( i\HT ( m\tilde{\omega}_2 + n\tilde{\omega}_3 )) 
= \sum_{\abs{m}+\abs{n} \leq 2\NT} \left( 1 - \frac{\max(|m|, |n|,|m-n|)}{2\NT+1} \right) \exp( i\HT ( m\tilde{\omega}_2 + n\tilde{\omega}_3 )) 
\end{align*}
where we have used the change of variables $m=j-l$, $n=k+l$, $o=l$. 
If $F_{2\NT}^{2D}$ denotes the corresponding convolution operator and 
$(D_{2\NT}^{2D}f)(\omega_1,\omega_2):=\frac{\HT^2}{(2\pi)^2}
\sum_{|m|+|n|\leq 2N}f(t_n,t_m)\exp(i\omega_1t_m+i\omega_2t_n)$
the two-dimensional orthogonal projection, we can use the inequality 
$\max(|m|, |n|,|m-n|)\leq |m|+|n|$ to obtain 
\begin{align}\label{eq:Fejer2Destim}
\begin{aligned}
\abs{\paren{D_{2\NT}^{2D}f-{F_{2\NT}^{2D}f}}
(\omega_1,\omega_2)}
&\leq \frac{\HT^2}{(2\pi)^2(2N+1)}\abs{\sum_{\abs{m}+\abs{n} \leq 2\NT} 
f(t_m,t_n)(|m|+|n|) e^{i\HT ( m\omega_1 + n\omega_2)}}
= \frac{1}{T} \abs{ \frac{\partial D_{2\NT}^{2D}f}{\partial \omega_1}(0,0)
+ \frac{\partial D_{2\NT}^{2D}f}{\partial \omega_2}(0,0)}.
\end{aligned}
\end{align}
If $f(\omega_1,\omega_2,\omega_3)$ is $2\pi/\HT$-periodic in all its arguments 
and $\tilde{f}(\tilde{\omega})=f(\omega(\tilde{\omega}))$, we find 
in analogy to section \ref{sect:lemma2} that 
\begin{align*}
\int d\tilde{\omega_2}\int d\tilde{\omega}_3 \
\mathcal{F}_{2\NT}^{2D}(3\tilde{\omega}_2,3\tilde{\omega}_3) 
\tilde{f}(\tilde{\omega}_1,\tilde{\omega}_2,\tilde{\omega}_3)
&= \frac{1}{9}\sum_{k,l=0}^{2}\paren{F_{2\NT}^{2D}\tilde{f}}
\paren{\tilde{\omega}_1,\frac{2\pi}{3\HT}k,\frac{2\pi}{3\HT}l}
= \frac{1}{9}\sum_{k,l=0}^{2}\paren{F_{2\NT}^{2D}\tilde{f}}
\paren{\tilde{\omega}_1-\frac{2\pi}{3\HT}(k+l),0,0},
\end{align*}
and hence  
$\int d\tilde{\omega_1}\int d\tilde{\omega_2}\int d\tilde{\omega}_3 \
\mathcal{F}_{2\NT}^{2D}(3\tilde{\omega}_2,3\tilde{\omega}_3) 
\tilde{f}(\tilde{\omega}_1,\tilde{\omega}_2,\tilde{\omega}_3)
= \int d\tilde{\omega_1} \ \paren{F_{2\NT}^{2D}\tilde{f}}
\paren{\tilde{\omega}_1,0,0}$. 
Together with \Eq{Fejer2Destim} we obtain 
\begin{equation}
 X = \frac{(2\pi)^2}{T^2} \int d\omega \ W_{1}(\omega) W_{2}(\omega) W_{3}(\omega) \Ecc_{12}(\omega)  \Ecc_{34}(\omega)  \Ecc_{56}(\omega) + \Mc{O}\left(\frac{1}{T^3} \right).
\end{equation}
The terms $Y_1$ is proved to be of very high order using the same method than in \Sec{lemma2}. The term $Y_2$ can be treated in the same way as it also contains a cosine that oscillates with $T$.  Finally, $Z$ is of order $1/T^3$ using a similar demonstration than in \Sec{lemma2}. 

\subsection{Functionals of nonseparable products of 
four $\Mcc_{ab}$ functions}\label{sect:lemma4}
Let $\Mcc$ be defined by \Eq{Mcc} and $W_i$ representing some functions of $\omega$. Then, we have the following expension:
\begin{align}
\begin{aligned}
 (2\pi)^4 \int d\omega_1 \int d\omega_2 \int d\omega_3 \int d\omega_4 & \ W_{1}(\omega_1) W_{2}(\omega_2) W_{3}(\omega_3)  W_{4}(\omega_4)  \Mcc_{12}(\omega_1, \omega_2)  \Mcc_{34}(\omega_1, \omega_3)  \Mcc_{56}(\omega_2, \omega_4) \Mcc_{78}(\omega_3, \omega_4) =  \\
 &\frac{(2\pi)^7}{T^3} \int d\omega \ W_{1}(\omega) W_{2}(\omega) W_{3}(\omega) W_{4}(\omega) \Ecc_{12}(\omega) \Ecc_{34}(\omega) \Ecc_{56}(\omega) \Ecc_{78}(\omega) + \Mc{O}\Bigl( \frac{1}{T^4} \Bigr). 
\end{aligned}\label{eq:lemma4}
\end{align}

As in the previous proof, different terms have to be treated. The terms with 
combinations of the expressions $I$ and $II$ can be bounded by the same 
methods as in \Sec{lemma3}, and the term involving only expressions $II$ can 
be bounded as in \Sec{lemma2}. The only different term is
\begin{equation}
 X := \int d\omega_1 \int d\omega_2 \int d\omega_3 \int d\omega_4 \ W_{1}(\omega_1) W_{2}(\omega_2) W_{3}(\omega_3) W_{3}(\omega_4) I_{12}(\omega_1, \omega_2) I_{34}(\omega_1, \omega_3)  I_{56}(\omega_2, \omega_4) I_{78}(\omega_3, \omega_4).
\end{equation}
Here small adaptions of the argument in \Sec{lemma3} with the change of variables
\begin{equation}\label{eq:COV_4d}
\int_{Q}d\omega \ f(\omega) = \int_Q d\tilde{\omega} \ f(\omega(\tilde{\omega})),\qquad 
\tilde{\omega} = \frac{1}{4}\left(\begin{smallmatrix}1 &1& 1& 1\\ -1& 1& 0& 0\\ -1& 0& 1& 0\\ -1& 0& 0& 1\end{smallmatrix}\right)\omega,
\qquad
\omega = \left(\begin{smallmatrix}1 &-1& -1& -1\\  1& 3& -1& -1\\ 1 &-1& 3& -1\\ 1 &-1 &-1 & 3\end{smallmatrix}\right)\tilde{\omega},
\qquad
Q:=[-\pi/\HT,\pi/\HT]^4.
\end{equation}
lead to the formula 
\begin{equation}
 X = \frac{(2\pi)^3}{T^3} \int d\omega \ W_{1}(\omega) W_{2}(\omega) W_{3}(\omega)  W_{4}(\omega) \Ecc_{12}(\omega) \Ecc_{34}(\omega) \Ecc_{56}(\omega) \Ecc_{78}(\omega) + \Mc{O}\Bigl( \frac{1}{T^4} \Bigr).
\end{equation}

\section{Noise covariance matrix for products of travel times} \label{sect:covtautauApp}
\subsection{Third order moment of the travel times} \label{sect:6thorder}
Using the definition of the travel times, we obtain that the covariance for the product of travel times is given by:
\begin{align*}
 \Cov[\tau_1(\xv_1, \xv_2) \tau_2(\xv_3, \xv_4),& \tau_3(\xv_5, \xv_6)] =  
  (2\pi)^3 \int d\omega_1 \int d\omega_2 \int d\omega_3 \ W_{12}^\ast(\omega_1) W_{34}^\ast(\omega_2) W_{56}(\omega_3) \times \\
&\times \Big \lbrace
  \Cov[C_{12}(\omega_1) C_{34}(\omega_2), C_{56}(\omega_3)] 
   - C^{\textrm{ref}}_{12}(\omega_1)\Cov[C_{34}(\omega_2), C_{56}(\omega_3)] - C^{\textrm{ref}}_{34}(\omega_2)\Cov[C_{12}(\omega_1), C_{56}(\omega_3)] \Big \rbrace.
\end{align*}
Using \Eq{covCCC} and the two results presented in \Secs{lemma2}{lemma3}, we can express the covariance for three travel-times as
\begin{align}
 \Cov[\tau_1(\xv_1, \xv_2) \tau_2(\xv_3, \xv_4), \tau_3(\xv_5, \xv_6)] &= \frac{(2\pi)^5}{T^2} \int d\omega \ W_{12}^\ast \Biggl( W_{34}^\ast \Bigl( W_{56} \bigl( \Ecc_{15} \Ecc_{32} \Ecc_{64} + \Ecc_{14} \Ecc_{62} \Ecc_{35} \bigr) + W_{56}^\ast \bigl( \Ecc_{14} \Ecc_{52} \Ecc_{36} + \Ecc_{16} \Ecc_{32} \Ecc_{54} \bigr) \Bigr)  \nonumber \\
 & \hspace{3cm} + W_{34} \Bigl( W_{56} \bigl( \Ecc_{15} \Ecc_{42} \Ecc_{63} + \Ecc_{13} \Ecc_{62} \Ecc_{45} \bigr) + W_{56}^\ast \bigl( \Ecc_{13} \Ecc_{52} \Ecc_{46} + \Ecc_{16} \Ecc_{42} \Ecc_{53} \bigr) \Bigr) \Biggr) \nonumber \\
 &-  \Et_1  \Cov[\tau_2(\xv_3, \xv_4), \tau_3(\xv_5, \xv_6)] - \Et_2 \Cov[\tau_1(\xv_1, \xv_2), \tau_3(\xv_5, \xv_6)] + \Mc{O}\left(\frac{1}{T^3}\right).  \label{eq:covtau3}
\end{align}
where $\Et_j$ is the expectation value of $\tau_j$ and the covariance involving two travel times can be computed with \Eq{covtau}.

\subsection{Analytic formula for the covariance matrix for products of travel times} \label{sect:8thorderFinal}

In this section, we derive the main result of this paper. It gives an analytic expression for the covariance matrix between a product of travel times.
Using the definition of the travel times, one can show that the covariance of the product of travel times is given by:
\begin{align}
\begin{aligned}
 \Cov[\tau_1 \tau_2, \tau_3 \tau_4] =& \;(2 \pi)^4 \int d\omega_1 \int d\omega_2 \int d\omega_3 \int d\omega_4  \ W_{12}^{\ast}(\omega_1) W_{34}^{\ast}(\omega_2) W_{56}(\omega_3) W_{78}(\omega_4) \times  \\
 & \Bigg\lbrace \Cov [ C_{12}(\omega_1) C_{34}(\omega_2), C_{56}(\omega_3) C_{78}(\omega_4) ]  \\
 & -C^{\textrm{ref}}_{78}(\omega_4) \Cov [C_{12}(\omega_1) C_{34}(\omega_2), C_{56}(\omega_3) ] 
  -C^{\textrm{ref}}_{56}(\omega_3) \Cov [C_{12}(\omega_1) C_{34}(\omega_2), C_{78}(\omega_4)] \\
 & -C^{\textrm{ref}}_{34}(\omega_2) \Cov [C_{12}(\omega_1), C_{56}(\omega_3) C_{78}(\omega_4)]
   -C^{\textrm{ref}}_{12}(\omega_1) \Cov [C_{34}(\omega_2), C_{56}(\omega_3) C_{78}(\omega_4)]  \\
  & + C^{\textrm{ref}}_{34}(\omega_2) \Big[ C^{\textrm{ref}}_{78}(\omega_4) \Cov [C_{12}(\omega_1), C_{56}(\omega_3)]
    + C^{\textrm{ref}}_{56}(\omega_3) \Cov [C_{12}(\omega_1), C_{78}(\omega_4)] \Big] \\
  & + C^{\textrm{ref}}_{12}(\omega_1) \Big[ C^{\textrm{ref}}_{78}(\omega_4) \Cov [C_{34}(\omega_2), C_{56}(\omega_3)]
     + C^{\textrm{ref}}_{56}(\omega_3) \Cov [C_{34}(\omega_2), C_{78}(\omega_4)] \Big] \Bigg\rbrace. 
\end{aligned}
\label{eq:covarianceMatrix}
\end{align}

 In \App{lemma}, we have shown that not all the terms will lead to the same number of frequency integrals. It implies that the covariance given by \Eq{covarianceMatrix} has terms of different order with respect to the observation time $T$. The terms containing 3 integrals in $\omega$ are of order $T^{-1}$ while the other ones are of order $T^{-2}$ and $T^{-3}$. We write the covariance as the sum between three terms for the different orders:
\begin{equation}
 \Cov[\tau_1 \tau_2, \tau_3 \tau_4] 
= \frac{1}{T} Z_1 + \frac{1}{T^2} Z_2 + \frac{1}{T^3} Z_3 
+ \Mc{O}\left(\frac{1}{T^4}\right). \label{eq:covarianceMatrix1}
\end{equation}
The terms of order $1/T^4$ come from the correlation between the frequencies in the frequency domain as detailed in \Sec{convergenceTauApp} for the covariance between travel times. The other terms are detailed below.

\subsubsection*{Term $Z_1$ of order $T^{-1}$}
Looking at \Eq{covarianceMatrix} one can see that this term is composed of
\begin{itemize}
 \item all the terms involving $\Cov[C, C]$,
 \item the terms with two integrals in $\omega$ for the terms with  $\Cov[C C, C]$ (term $\Lambda^3_2$),
 \item the terms with three integrals in $\omega$ for the terms with  $\Cov[C C, C C]$ (term $\Lambda^4_3$)
\end{itemize}
where $C$ is a generic cross-covariance. Reorganizing terms leads to the formula \Eq{covtautau1} for $Z_1$. 

\subsubsection*{Term $Z_2$ of order $T^{-2}$}
Looking at \Eq{covarianceMatrix} one can see that this term is composed of
\begin{itemize}
 \item the terms with one integral in $\omega$ for the terms with  $\Cov[C C, C]$ (term $\Lambda^3_1$)
 \item the terms with two integrals in $\omega$ for the terms with  $\Cov[C C, C C]$ (term $\Lambda^4_2$).
\end{itemize}
Reorganizing terms leads to the formula \Eq{Z2} for $Z_2$. 

\subsubsection*{Term $Z_3$ of order $T^{-3}$}
The terms of order $T^{-3}$ come for the terms with only one integral in $\omega$ in $\Cov[C C, C C]$ (term $\Lambda^4_1$). 
This yields \Eq{Z3text} for $Z_3$. 

\section{Far-field approximation for $\Var[\tau^2_\textrm{diff}(\Delta)]$} \label{sect:farFieldApp}

In this section we give approximate expressions for the different terms composing \Eq{covarianceMatrixText} for $\Var[\tau^2_\textrm{diff}(\Delta)]$ in the far field ($\Delta \rightarrow \infty$). We start with the definitions of $Z_1$, $Z_2$ and $Z_3$:
\begin{align*}
 \frac{1}{T}Z_1 &= 4 \Et(\Delta)^2 \Var[\tau(\Delta)], \\
 \frac{1}{T^2}Z_2 &= 2 (\Var[\tau(\Delta)])^2 - 4 \Et(\Delta) \frac{(2\pi)^5}{T^2} \int d\omega \ \abs{W(\Delta, \omega)}^2 \Ecc(\Delta, \omega) \Ecc(0, \omega) \times \Bigl(W(\Delta, \omega)\Ecc(0, \omega) + W^\ast(\Delta, \omega)\Ecc(\Delta, \omega) \Bigr), \\
 \frac{1}{T^3}Z_3 &= 3 \frac{(2\pi)^7}{T^3} \int d\omega \ \abs{W(\Delta, \omega)}^2 \Bigl( W(\Delta, \omega)\Ecc(0, \omega)^2 + W^\ast(\Delta, \omega)\Ecc(\Delta, \omega)^2 \Bigr)^2.
\end{align*}
In the far field, we have $\Ecc(\Delta, \omega) \ll \Ecc(0, \omega)$. If we suppose that $C^{\textrm{ref}} = (1 + \epsilon) \Ecc$ then the global behaviour of the four terms is:
\begin{align}
 \frac{1}{T}Z_1 &\sim 4 (2\pi)^3 \frac{\epsilon^2}{T}  \left( \int d\omega \ W^\ast(\Delta, \omega) \Ecc(\Delta, \omega)  \right)^2 \int d\omega \abs{W(\Delta, \omega)}^2 \Ecc(0, \omega)^2  \nonumber \\
 \frac{1}{T^2}Z_2 & \sim  2 \frac{(2\pi)^6}{T^2}  \left( \int d\omega \ \abs{W(\Delta, \omega)}^2  \Ecc(0, \omega)^2 \right)^2 + 4 (2\pi)^5 \frac{\epsilon}{T^2}  \left( \int d\omega \ \abs{W(\Delta, \omega)}^2 W(\Delta, \omega) \Ecc(0, \omega)^2 \Ecc(\Delta, \omega) \right) \times \int d\omega \ W^\ast(\Delta, \omega) \Ecc(\Delta, \omega)  \nonumber \\
 \frac{1}{T^3}Z_3 & \sim 3 \frac{(2 \pi)^7}{T^3} \int d\omega  \ \abs{W(\Delta, \omega)}^4 \Ecc(0, \omega)^4. \label{eq:comparisonFarfield}
\end{align}
We can thus see that the global behaviour of the terms is
\begin{align*}
 \frac{1}{T}Z_1 & \sim \frac{\epsilon^2}{T} \Ecc(\Delta, \omega)^2 \Ecc(0, \omega)^2  \\
 \frac{1}{T^2}Z_2 & \sim  \frac{1}{T^2} \Ecc(0, \omega)^4 +  \frac{\epsilon}{T^2} \Ecc(0, \omega)^3 \Ecc(\Delta, \omega) \\
 \frac{1}{T^3}Z_3 & \sim \frac{1}{T^3} \Ecc(0, \omega)^4.
\end{align*}
As $\Ecc(\Delta, \omega) \ll \Ecc(0, \omega)$ we can conclude that in this case the first term in $Z_2$ and the one in $Z_3$ are dominant. We can go further to see for which observation time $T_c$ these two last terms intersect in the case of difference travel times. If the window function $f(t)$ in the definition of $W_{\textrm{diff}}$ defined by \Eq{W} is a Heavyside function then we have \citep{GIZ04}
\begin{equation}
 W_{\textrm{diff}}(\Delta, \omega) = \frac{2i\omega C^{\textrm{ref}}(\Delta, \omega)^\ast}{2\pi h_\omega \sum_{\omega'} {\omega'}^2 \abs{C^{\textrm{ref}}(\Delta, \omega')}^2}.
\end{equation}
For a $p-$mode ridge $\kappa_r = \kappa_r(\omega)$ the function $\Ecc(\Delta, \omega)$ can be written in the far field as \citep{GIZ04}
\begin{equation}
 \Ecc(\Delta, \omega) \approx \sqrt{\frac{2}{\pi \kappa_r \Delta}} \Ecc(0, \omega) \textrm{e}^{-\kappa_i \Delta} \cos \left( \kappa_r \Delta - \frac{\pi}{4} \right) \label{eq:CdeltaFarField}
\end{equation}
where $\kappa_i$ us the imaginary part of the wavenumber at resonance and represents attenuation of the waves. The sums in \Eq{comparisonFarfield} can be approximated using the fact that the cosine in \Eq{CdeltaFarField} oscillates many times within the frequency width $\xi$ of the envelope of $\Ecc(0,\omega)$ such that
\begin{equation*}
\frac{1}{T^2}Z_2 \approx 2 \frac{(2\pi)^6}{T^2} \left( \frac{2 \pi \kappa_r \Delta \textrm{e}^{2\kappa_i \Delta}}{ \xi \omega_0^2} \right)^2 \quad \textrm{ and } \quad \frac{1}{T^3}Z_3 \approx 3 \frac{(2 \pi)^7}{T^3} \frac{\kappa_r^2 \Delta^2 \textrm{e}^{4\kappa_i \Delta}}{\pi^2 \omega_0^4 \xi^3}. 
\end{equation*}
Using the numerical value $\xi / 2\pi = 1$mHz, the observation time $T_c$ at which the two terms are equal is
\begin{equation}
 T_c = T \frac{Z_2}{Z_3} \approx \frac{12 \pi}{\xi}  = 100 \textrm{min}.
\end{equation}
For $T > T_c$, $Z_2 / T^2$ is the dominant term. As the observation time is traditionally of at least eight hours in helioseismology, the term of order $1/T^3$ can be neglected.

\end{document}